\documentstyle[prd,aps,epsf,floats,amsfonts,amssymb,amsmath]{revtex}
\newcommand{\be}{\begin{equation}}\newcommand{\ee}{\end{equation}}
\newcommand{\bea}{\begin{eqnarray}}\newcommand{\eea}{\end{eqnarray}}
\newcommand{\brr}{\begin{array}}\newcommand{\err}{\end{array}}
\newcommand{\bit}{\begin{itemize}}\newcommand{\eit}{\end{itemize}}
\newcommand{\ben}{\begin{enumerate}}\newcommand{\een}{\end{enumerate}}

\def\lab{\label}\def\lan{\langle}
\def\lf{\left}\def\lrar{\leftrightarrow}
\def\noi{\noindent}\def\non{\nonumber}
\def\pa{\partial}\def\ran{\rangle}\def\rar{\rightarrow}
\def\Rar{\Rightarrow}\def\ri{\right}\def\ti{\tilde}
\def\wti{\widetilde}
\def\al{\alpha}\def\bt{\beta}\def\ga{\gamma}
\def\de{\delta}\def\De{\Delta}
\def\te{\theta}\def\la{\lambda}
\def\si{\sigma}\def\om{\omega}
\def\Om{\Omega}

\def\CP{{_{C\!P}}}\def\T{{_{T}}}

\def\1{{_{1}}}\def\2{{_{2}}}
\begin{document}

\title{Quantum Field Theory of three flavor neutrino mixing
\\ and oscillations with  CP violation}
\author{Massimo Blasone${}^{\sharp\flat}$, Antonio Capolupo${}^{\flat}$
and Giuseppe Vitiello${}^{\flat }$ \vspace{3mm}}

\address{${}^{\sharp}$
Blackett  Laboratory, Imperial  College,   Prince Consort Road,
London  SW7 2BW, U.K.
\\ [2mm] ${}^{\flat}$
Dipartimento di  Fisica and INFN, Universit\`a di Salerno, I-84100
Salerno, Italy \vspace{2mm}}


\maketitle

\begin{abstract}
We study in detail the Quantum Field Theory of mixing among three
generations of Dirac fermions (neutrinos). We construct the
Hilbert space for the flavor fields and determine the generators
of the mixing transformations. By use of these generators, we
recover all the known parameterizations of the three-flavor mixing
matrix and we find a number of new ones. The algebra of the
currents associated with the mixing transformations is shown to be
a deformed $su(3)$ algebra, when CP violating phases are present.
We then derive the exact oscillation formulas, exhibiting new
features with respect to the usual ones. CP and T violation are
also discussed.
\end{abstract}

\vspace{8mm}


\section{Introduction}
In recent years, there has been much progress in the understanding
of flavor mixing in Quantum Field Theory (QFT)
\cite{BV95,BHV99,remarks,currents,hannabuss,fujii1,fujii2,comment,lathuile,binger,bosonmix,Ji2,Ji3,Beuthe,Grimus,Cardall,Muller}.
The original discovery of the unitary inequivalence of the mass
and the flavor representations in QFT \cite{BV95}, has prompted
further investigations on fermion mixing
\cite{BHV99,remarks,currents,hannabuss,fujii1,fujii2,comment} as
well as on boson
mixing\cite{comment,lathuile,binger,bosonmix,Ji2,Ji3}. It has
emerged that the rich non--perturbative vacuum structure
associated with field mixing leads to relevant modification of the
flavor oscillation formulas, exhibiting new features with respect
to the usual quantum--mechanical ones \cite{Pontec}. Some
topologically nontrivial features, such as the occurrence of a
geometric (Berry-Anandan) phase in field mixing has been also
pointed out \cite{Berry}.

In this paper we study in detail the case of three flavor fermion
(neutrino) mixing. This is not a simple extension of the previous
results\cite{BV95,BHV99,remarks} since the existence of a CP
violating phase in the parameterization of the three-flavor mixing
matrix introduces novel features which are absent in the
two-flavor case. We determine the generators of the mixing
transformations and by use of them, we recover the known
parameterizations of the three-flavor mixing matrix and  find a
number of new ones. We construct the flavor Hilbert space, for
which the ground state (flavor vacuum) turns out to be a
generalized coherent state. We also study the algebraic structure
of currents and charges associated with the mixing transformations
and we find that, in presence of CP violation, it is that of a
deformed $su(3)$. The construction of the flavor Hilbert space is
an essential step in the derivation of exact oscillation formulas,
which account for CP violation and reduce to the corresponding
quantum--mechanical ones in the relativistic limit.

The paper is organized as follows. In \S II  we construct the
Hilbert space for three-flavor mixed fermions. In \S III we study
the various parameterization of the unitary $3\times3$ mixing
matrix obtained by use of the algebraic generators. In \S IV, we
study the currents and charges for three-flavor mixing, which are
then used in \S V  to derive the exact neutrino oscillation
formulas. Finally in \S  VI, CP an T violation in QFT neutrino
oscillations are discussed. \S VII is devoted to conclusions. In
the Appendices we put some useful formulas and a discussion of the
arbitrary mass parameterization in the expansion of flavor fields
as recently reported in \cite{fujii2,comment}.

\section{Three flavor fermion mixing}

We start by considering the following Lagrangian density
describing three Dirac fields with a mixed mass term:
\bea\label{lagemu} {\cal L}(x)\,=\,  {\bar \Psi_f}(x) \lf( i
\not\!\partial - \textsf{M} \ri) \Psi_f(x)\, , \eea
where $\Psi_f^T=(\nu_e,\nu_\mu,\nu_{\tau})$ and $\textsf{M} =
\textsf{M}^\dag$ is the mixed mass matrix.

Among the various possible parameterizations of the mixing matrix
for three fields, we choose to work with the following one since
it is the familiar parametrization of the CKM matrix
\cite{Cheng-Li,CKM}:
\bea\label{fermix} \Psi_f(x) \, = {\cal U} \, \Psi_m
(x)=\begin{pmatrix}
c_{12}c_{13} & s_{12}c_{13} & s_{13}e^{-i\de} \\
-s_{12}c_{23}-c_{12}s_{23}s_{13}e^{i\de} &
c_{12}c_{23}-s_{12}s_{23}s_{13}e^{i\de} & s_{23}c_{13} \\
s_{12}s_{23}-c_{12}c_{23}s_{13}e^{i\de} &
-c_{12}s_{23}-s_{12}c_{23}s_{13}e^{i\de} & c_{23}c_{13}
\end{pmatrix}\,\Psi_m (x) \, , \eea
with $c_{ij}=\cos\te_{ij}$ and  $s_{ij}=\sin\te_{ij}$, being
$\te_{ij}$ the mixing angle between $\nu_{i},\nu_{j}$ and
$\Psi_m^T=(\nu_1,\nu_2,\nu_3)$.

Using Eq.(\ref{fermix}), we diagonalize the quadratic form of
Eq.(\ref{lagemu}), which then reduces to the Lagrangian for three
Dirac fields, with masses $m_1$, $m_2$ and $m_3$:
\bea\label{lag12} {\cal L}(x)\,=\,  {\bar \Psi_m}(x) \lf( i
\not\!\partial -  \textsf{M}_d\ri) \Psi_m(x)  \, , \eea
where $\textsf{M}_d = diag(m_1,m_2,m_3)$.

Following Ref.\cite{BV95}, we construct the generator for the
mixing transformation (\ref{fermix}) and define\footnote{Let us
consider for example the generation of the first row of the mixing
matrix ${\cal U}$. We have (see also Appendix B) $ \pa \nu_e
/\pa\te_{23}\, =\, 0; $ and $ \pa\nu_e /\pa \te_{13} \, =
  \,G_{12}^{-1}G_{13}^{-1}[\nu_1,L_{13}]G_{13}G_{12}\,=\,
  G_{12}^{-1}G_{13}^{-1} e^{-i \de}\nu_3 G_{13}G_{12} $, thus:
\bea
 \non \pa^2\nu_e/\pa \te_{13}^2  \, = \, -\nu_e \quad \Rar \quad
\nu_e \, = \, f(\te_{12}) \cos\te_{13} + g(\te_{12}) \sin\te_{13};
\eea
with the initial conditions (from Eq.(\ref{incond})): $f(\te_{12})
= \nu_e|_{\te_{13}=0}  $ and $ g(\te_{12}) = \pa\nu_e/\pa
\te_{13}|_{\te_{13}=0} = e^{-i \de}\nu_3 $ . We also have
\bea \non \pa^2f(\te_{12})/\pa \te_{13}^2  \,  = \, -f(\te_{12})
\quad \Rar \quad f(\te_{12}) \, = \, A \cos\te_{12}  \,+ B
\sin\te_{12} \eea
with the initial conditions $A=\nu_e|_{\te=0}=\nu_1$ and $B=\pa
f(\te_{12}) /\pa\te_{12}|_{\te=0}=\nu_2$, and $\te=(\te_{12},
\te_{13}, \te_{23}).$}
\bea\label{incond} &&\nu_{\si}^{\al}(x)\equiv G^{-1}_{\bf \te}(t)
\, \nu_{i}^{\al}(x)\, G_{\bf \te}(t), \eea where $(\si,i)=(e,1),
(\mu,2), (\tau,3)$, and
\bea\label{generator} &&G_{\bf
\te}(t)=G_{23}(t)G_{13}(t)G_{12}(t)\, , \eea
where
\bea\label{generators1} && G_{12}(t)\equiv
\exp\Big[\te_{12}L_{12}(t)\Big]\;\;\;\;\;\;;\;\;\;\;
L_{12}(t)=\int
d^{3}x\lf[\nu_{1}^{\dag}(x)\nu_{2}(x)-\nu_{2}^{\dag}(x)\nu_{1}(x)\ri],
\\ \label{generators2}
&&G_{23}(t)\equiv\exp\Big[\te_{23}L_{23}(t)\Big]\;\;\;\;\;\;;\;\;\;\;
L_{23}(t)=\int
d^{3}x\lf[\nu_{2}^{\dag}(x)\nu_{3}(x)-\nu_{3}^{\dag}(x)\nu_{2}(x)\ri],
\\ \label{generators3}
&&G_{13}(t)\equiv\exp\Big[\te_{13}L_{13}(\de,t)\Big]
\;\;\;\;;\;\;\;\;L_{13}(\de,t)=\int
d^{3}x\lf[\nu_{1}^{\dag}(x)\nu_{3}(x)e^{-i\de}-\nu_{3}^{\dag}(x)
\nu_{1}(x)e^{i\de}\ri], \eea
It is evident from the above form of the generators, that the
phase $\de$ is unavoidable for three field mixing, while it can be
incorporated in the definition of the fields in the two flavor
case.

The free fields  $\nu_i$ (i=1,2,3) can be quantized in the usual
way\cite{Itz} (we use $t\equiv x_0$):
\bea\label{2.2} \nu_{i}(x) = \sum_{r} \int d^3 k \lf[u^{r}_{{\bf
k},i}(t) \al^{r}_{{\bf k},i}\:+    v^{r}_{-{\bf k},i}(t)
\bt^{r\dag }_{-{\bf k},i}  \ri] e^{i {\bf k}\cdot{\bf x}} ,\qquad
i=1,2,3\,, \eea
with $u^{r}_{{\bf k},i}(t)=e^{-i\om_{k,i} t}u^{r}_{{\bf k},i}$,
$v^{r}_{{\bf k},i}(t)=e^{i\om_{k,i} t}v^{r}_{{\bf k},i}$ and
$\om_{k,i}=\sqrt{{\bf k}^2+m_i^2}$. The vacuum for the mass
eigenstates is denoted by $|0\ran_{m}$:  $\; \; \al^{r}_{{\bf
k},i}|0\ran_{m}= \bt^{r }_{{\bf k},i}|0\ran_{m}=0$.   The
anticommutation relations are the usual ones; the wave function
orthonormality and completeness relations are those of
Ref.\cite{BV95}.

The main result of Ref.\cite{BV95} is the unitary inequivalence
(in the infinite volume limit) of the vacua for the flavor fields
and for the fields with definite masses. There such an
inequivalence was proved for the case of two generations;
subsequently, in Ref.\cite{hannabuss}, a rigorous general proof of
such inequivalence for any number of generations has been given
(see also Ref.\cite{Ji3}). Thus we do not need to repeat here such
a proof and we define the the {\em flavor vacuum} as:
\bea\label{flavac}
|0(t)\ran_{f}\,\equiv\,G_{\te}^{-1}(t)\;|0\ran_{m} \;. \eea

The form of this state is considerably more complicated of the one
for two generations. When $\de=0$, the generator $G_\te$ is an
element of the $SU(3)$ group (see \S IV) and  the flavor vacuum is
classified as an $SU(3)$ generalized coherent state \`a la
Perelomov \cite{perelomov}. A nonzero CP violating phase
introduces an interesting modification of the algebra associated
with the mixing transformations Eq.(\ref{fermix}): we discuss this
in \S IV.

By use of $G_{\bf \te}(t)$, the flavor fields can be expanded as:
\bea\label{exnue1} &&{}\quad\qquad \nu_\si(x)= \sum_{r} \int d^3 k
\lf[ u^{r}_{{\bf k},i}(t) \al^{r}_{{\bf k},\si}(t) +
v^{r}_{-{\bf k},i}(t) \bt^{r\dag}_{-{\bf k},\si}(t) \ri]  e^{i
{\bf k}\cdot{\bf x}}\quad,\quad (\si,i)=(e,1) , (\mu,2) ,
(\tau,3)\, . \eea

The flavor annihilation operators are defined as $\al^{r}_{{\bf
k},\si}(t) \equiv G^{-1}_{\bf \te}(t)\al^{r}_{{\bf k},i} G_{\bf
\te}(t)$ and $\bt^{r\dag}_{{-\bf k},\si}(t)\equiv
 G^{-1}_{\bf \te}(t) \bt^{r\dag}_{{-\bf k},i}
G_{\bf \te}(t)$. They clearly act as annihilators for the flavor
vacuum Eq.(\ref{flavac}). For further reference, it is useful to
list explicitly the flavor annihilation/creation operators (see
also Ref.\cite{BV95}). In the reference frame ${\bf k}=(0,0,|{\bf
k}|)$ the spins decouple and their form is particularly simple:
\bea \al_{{\bf k},e}^{r}(t)&=&c_{12}c_{13}\;\al_{{\bf k},1}^{r} +
s_{12}c_{13}\lf(U^{{\bf k}*}_{12}(t)\;\al_{{\bf k},2}^{r}
+\epsilon^{r} V^{{\bf k}}_{12}(t)\;\bt_{-{\bf k},2}^{r\dag}\ri) +
e^{-i\de}\;s_{13}\lf(U^{{\bf k}*}_{13}(t)\;\al_{{\bf k},3}^{r}
+\epsilon^{r} V^{{\bf k}}_{13}(t)\;\bt_{-{\bf k},3}^{r\dag}\ri)\;,
\\[2mm]\non
\al_{{\bf k},\mu}^{r}(t)&=&\lf(c_{12}c_{23}- e^{i\de}
\;s_{12}s_{23}s_{13}\ri)\;\al_{{\bf k},2}^{r} -
\lf(s_{12}c_{23}+e^{i\de}\;c_{12}s_{23}s_{13}\ri) \lf(U^{{\bf
k}}_{12}(t)\;\al_{{\bf k},1}^{r} -\epsilon^{r} V^{{\bf
k}}_{12}(t)\;\bt_{-{\bf k},1}^{r\dag}\ri)
\\
&&+\;s_{23}c_{13}\lf(U^{{\bf k}*}_{23}(t)\;\al_{{\bf k},3}^{r} +
\epsilon^{r} V^{{\bf k}}_{23}(t)\;\bt_{-{\bf k},3}^{r\dag}\ri)\;,
\\[2mm]\non
\al_{{\bf k},\tau}^{r}(t)&=&c_{23}c_{13}\;\al_{{\bf k},3}^{r} -
\lf(c_{12}s_{23}+e^{i\de}\;s_{12}c_{23}s_{13}\ri) \lf(U^{{\bf
k}}_{23}(t)\;\al_{{\bf k},2}^{r} -\epsilon^{r} V^{{\bf
k}}_{23}(t)\;\bt_{-{\bf k},2}^{r\dag}\ri)
\\
&&+\;\lf(s_{12}s_{23}- e^{i\de}\;c_{12}c_{23}s_{13}\ri)
\lf(U^{{\bf k}}_{13}(t)\;\al_{{\bf k},1}^{r} -\epsilon^{r} V^{{\bf
k}}_{13}(t)\;\bt_{-{\bf k},1}^{r\dag}\ri)\;, \eea

\bea \bt^{r}_{-{\bf k},e}(t)&=&c_{12}c_{13}\;\bt_{-{\bf k},1}^{r}
+ s_{12}c_{13}\lf(U^{{\bf k}*}_{12}(t)\;\bt_{-{\bf k},2}^{r}
-\epsilon^{r}V^{{\bf k}}_{12}(t)\;\al_{{\bf k},2}^{r\dag}\ri)
+e^{i\de}\; s_{13}\lf(U^{{\bf k}*}_{13}(t)\;\bt_{-{\bf k},3}^{r}
-\epsilon^{r} V_{13}^{{\bf k}}(t)\;\al_{{\bf k},3}^{r\dag}\ri)\;,
\\[2mm] \non
\bt^{r}_{-{\bf k},\mu}(t)&=&\lf(c_{12}c_{23}- e^{-i\de}\;
s_{12}s_{23}s_{13}\ri)\;\bt_{-{\bf k},2}^{r} -
\lf(s_{12}c_{23}+e^{-i\de}\;c_{12}s_{23}s_{13}\ri) \lf(U^{{\bf
k}}_{12}(t)\;\bt_{-{\bf k},1}^{r} +\epsilon^{r}\; V^{{\bf
k}}_{12}(t)\;\al_{{\bf k},1}^{r\dag}\ri) +
\\
&&+\; s_{23}c_{13}\lf(U^{{\bf k}*}_{23}(t)\;\bt_{-{\bf k},3}^{r} -
\epsilon^{r}\; V^{{\bf k}}_{23}(t)\;\al_{{\bf k},3}^{r\dag}\ri)\;,
\\[2mm] \non
\bt^{r}_{-{\bf k},\tau}(t)&=&c_{23}c_{13}\;\bt_{-{\bf k},3}^{r} -
\lf(c_{12}s_{23}+e^{-i\de}\;s_{12}c_{23}s_{13}\ri) \lf(U^{{\bf
k}}_{23}(t)\;\bt_{-{\bf k},2}^{r} + \epsilon^{r} V^{{\bf
k}}_{23}(t)\;\al_{{\bf k},2}^{r\dag}\ri)
\\
&&+\;\lf(s_{12}s_{23}- e^{-i\de}\;c_{12}c_{23}s_{13}\ri)
\lf(U^{{\bf k}}_{13}(t)\;\bt_{-{\bf k},1}^{r} + \epsilon^{r}
V^{{\bf k}}_{13}(t)\;\al_{{\bf k},1}^{r\dag}\ri)\;. \eea

These operators satisfy canonical (anti)commutation relations at
equal times. The main difference with respect to their ``naive''
quantum-mechanical counterparts is in the anomalous terms
proportional to the $V_{ij}$ factors. In fact, $U^{{\bf k}}_{ij}$
and $V^{{\bf k}}_{ij}$ are Bogoliubov coefficients defined as:
\bea V^{{\bf k}}_{ij}(t)=|V^{{\bf
k}}_{ij}|\;e^{i(\om_{k,j}+\om_{k,i})t}\;\;\;\;,\;\;\;\; U^{{\bf
k}}_{ij}(t)=|U^{{\bf k}}_{ij}|\;e^{i(\om_{k,j}-\om_{k,i})t} \eea

\bea &&|U^{{\bf
k}}_{ij}|=\lf(\frac{\om_{k,i}+m_{i}}{2\om_{k,i}}\ri)
^{\frac{1}{2}}
\lf(\frac{\om_{k,j}+m_{j}}{2\om_{k,j}}\ri)^{\frac{1}{2}}
\lf(1+\frac{|{\bf k}|^{2}}{(\om_{k,i}+m_{i})
(\om_{k,j}+m_{j})}\ri)=\cos(\xi_{ij}^{{\bf k}})
\\
&&|V^{{\bf k}}_{ij}|=\lf(\frac{\om_{k,i}+m_{i}}{2\om_{k,i}}\ri)
^{\frac{1}{2}}
\lf(\frac{\om_{k,j}+m_{j}}{2\om_{k,j}}\ri)^{\frac{1}{2}}
\lf(\frac{|{\bf k}|}{(\om_{k,j}+m_{j})}-\frac{|{\bf
k}|}{(\om_{k,i}+m_{i})}\ri)=\sin(\xi_{ij}^{{\bf k}}) \eea \bea
|U^{{\bf k}}_{ij}|^{2}+|V^{{\bf k}}_{ij}|^{2}=1 \eea where
$i,j=1,2,3$ and $j>i$.
The following identities hold:

\bea \label{ident1} &&V^{{\bf k}}_{23}(t)V^{{\bf
k}*}_{13}(t)+U^{{\bf k}*}_{23}(t)U^{{\bf k}}_{13}(t) = U^{{\bf
k}}_{12}(t)\;\;\;\;\;,\;\;\;\;\; V^{{\bf k}}_{23}(t)U^{{\bf
k}*}_{13}(t)-U^{{\bf k}*}_{23}(t)V^{{\bf k}}_{13}(t) =- V^{{\bf
k}}_{12}(t)
\\ \label{ident2}
&&U^{{\bf k}}_{12}(t)U^{{\bf k}}_{23}(t)-V^{{\bf
k}*}_{12}(t)V^{{\bf k}}_{23}(t) = U^{{\bf
k}}_{13}(t)\;\;\;\;\;\;,\;\;\;\;\; U^{{\bf k}}_{23}(t)V^{{\bf
k}}_{12}(t)+U^{{\bf k}*}_{12}(t)V^{{\bf k}}_{23}(t) = V^{{\bf
k}}_{13}(t)
\\ \label{ident3}
&&V^{{\bf k}*}_{12}(t)V^{{\bf k}}_{13}(t)+U^{{\bf
k}*}_{12}(t)U^{{\bf k}}_{13}(t) = U^{{\bf
k}}_{23}(t)\;\;\;\;\;,\;\;\;\;\; V^{{\bf k}}_{12}(t)U^{{\bf
k}}_{13}(t)-U^{{\bf k}}_{12}(t)V^{{\bf k}}_{13}(t) =- V^{{\bf
k}}_{23}(t)\;,
\\ \label{ident4}
&& \xi_{13}^{{\bf k}}=\xi_{12}^{{\bf k}} +\xi_{23}^{{\bf k}}
\qquad, \qquad
 \xi_{ij}^{{\bf k}}= \arctan\lf(|V^{{\bf k}}_{ij}|\,/\,|U^{{\bf k}}_{ij}|\ri)\,.
\eea

As already observed in Ref.\cite{BV95} we remark that, in contrast
with the case of two flavor mixing, the condensation densities are
now different for particles of different masses:
\bea {\cal N}^{\bf k}_1\, = \,_{f}\langle0(t)|N^{{\bf
k},r}_{\al_{1}} |0(t)\ran_{f}&=& \,_{f}\langle0(t)|N^{{\bf
k},r}_{\bt_{1}}|0(t)\ran_{f}= s^{2}_{12}c^{2}_{13}\,|V^{{\bf
k}}_{12}|^{2}+ s^{2}_{13}\,|V^{{\bf k}}_{13}|^{2}\,,
\\
{\cal N}^{\bf k}_2\, = \,_{f}\langle0(t)|N^{{\bf
k},r}_{\al_{2}}|0(t)\ran_{f}&=& \,_{f}\langle0(t)|N^{{\bf
k},r}_{\bt_{2}}|0(t)\ran_{f}
=\lf|-s_{12}c_{23}+e^{i\de}\,c_{12}s_{23}s_{13}\ri|^{2} \,|V^{{\bf
k}}_{12}|^{2}+ s^{2}_{23}c^{2}_{13}\;|V^{{\bf k}}_{23}|^{2}\,,
\\
{\cal N}^{\bf k}_3\, = \,_{f}\langle0(t)|N^{{\bf
k},r}_{\al_{3}}|0(t)\ran_{f}&=& \,_{f}\langle0(t)|N^{{\bf
k},r}_{\bt_{3}}|0(t)\ran_{f}=
\lf|-c_{12}s_{23}+e^{i\de}\,s_{12}c_{23}s_{13}\ri|^{2} |V^{{\bf
k}}_{23}|^{2} + \lf|s_{12}s_{23}+
e^{i\de}\,c_{12}c_{23}s_{13}\ri|^{2} |V^{{\bf k}}_{13}|^{2}\,.
\eea

\vspace{0.4cm}

$$
\begin{tabular}{|c|c|c|c|c|c|c|}
\hline \;\;{\em $m_{1}^{}$} \; & \;\;{\em $m_{2}$}\; & \;\;{\em
$m_{3}$}\; & \;\;{\em $\te_{12}$}\; &\;\;{\em $\te_{13}$} \;&
\;\;{\em $\te_{23}$}\;&
\;\;{\em $\de$}\; \\[1.5mm] \hline
$ 1^{}$ & $ 200$   &  $ 3000 $    &  $  \pi/4  $  &  $ \pi/4  $ &
$ \pi/4  $&  $\; \pi/4  $
\\[1.5mm] \hline
\end{tabular}
$$
\centerline{\small Table 1: The values of masses and mixing angles
used for plots }

\vspace{0.5cm} We plot in Fig.(1) the condensation densities for
sample values of parameters as given in the table
above\footnote{Here and in the following plots,
 we use the same (energy) units for the
values of masses and momentum.}: \vspace{0.5cm}

\centerline{\epsfysize=3.0truein\epsfbox{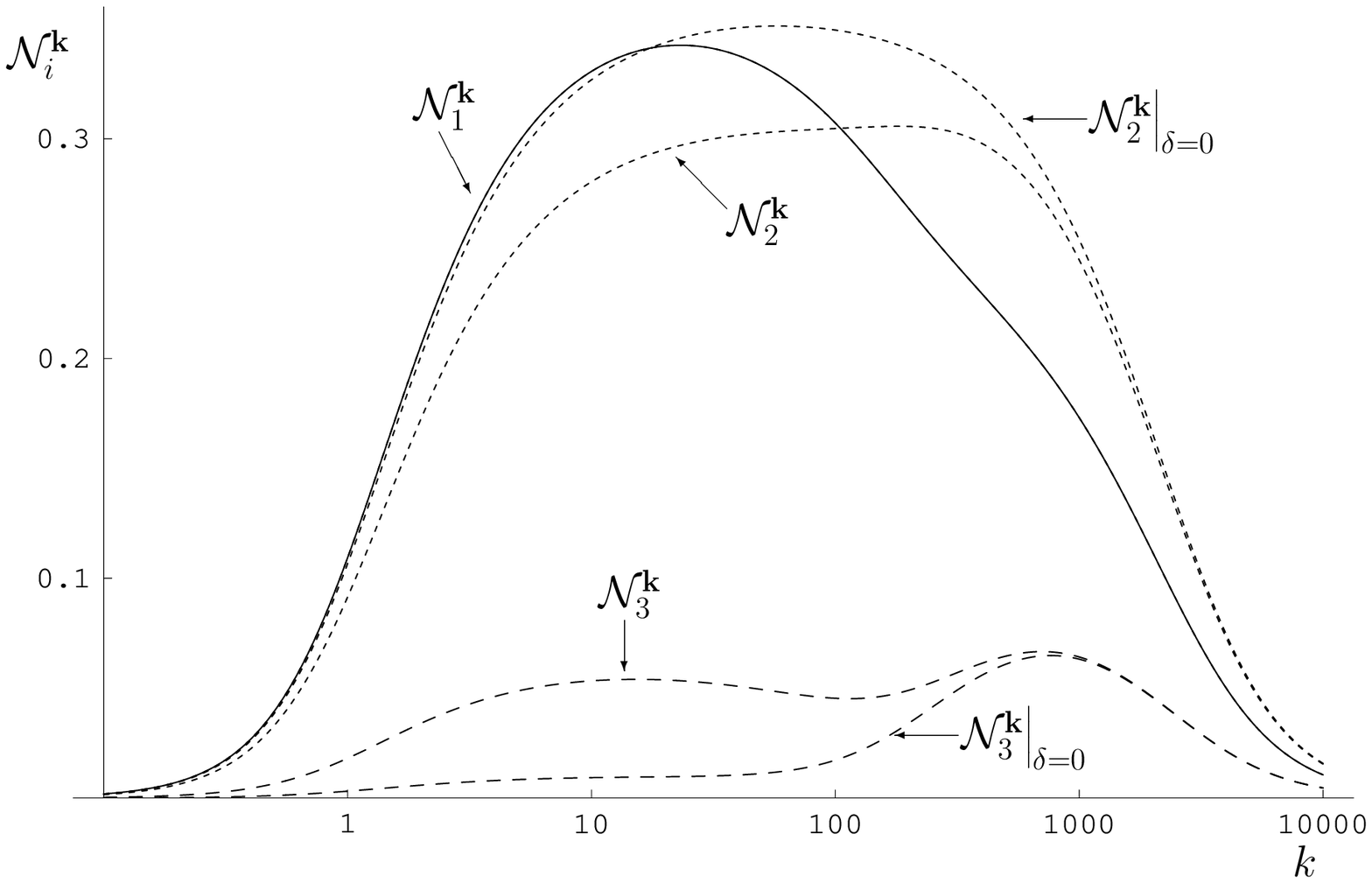}} \vspace{.2cm}
\centerline{\small Figure 1:  Plot of the condensation densities
${\cal N}^{\bf k}_i$ in function of $|{\bf k}|$  for the values of
parameters as in Tab.(1). }

\vspace{0.5cm}

\section{The  parameterizations of the three flavor mixing
matrix}

In \S II we have studied the generator of the mixing matrix ${\cal
U}$ of Eq.(\ref{fermix}). However, this matrix is only one of the
various forms in which a $3\times 3$ unitary matrix can be
parameterized. Indeed, the generator Eq.(\ref{generator}) can be
used for generating such alternative parameterizations. To see
this, let us first  define in a more general way the generators
$G_{ij}$ including phases for all of them:

\bea\ &&G_{12}(t)\equiv\exp\Big[\te_{12}L_{12}(\de_{12},t)\Big]
\;\;\;;\;\;\; L_{12}(\de_{12},t)=\int
d^{3}x\lf[\nu_{1}^{\dag}(x)\nu_{2}(x)e^{-i\de_{12}}-
\nu_{2}^{\dag}(x)\nu_{1}(x)e^{i\de_{12}}\ri],
\\
&&G_{23}(t)\equiv \exp\Big[\te_{23}L_{23}(\de_{23},t)\Big]
\;\;\;;\;\;\; L_{23}(\de_{23},t)=\int
d^{3}x\lf[\nu_{2}^{\dag}(x)\nu_{3}(x)e^{-i\de_{23}}-
\nu_{3}^{\dag}(x)\nu_{2}(x)e^{i\de_{23}}\ri],
\\
&&G_{13}(t)\equiv\exp\Big[\te_{13}L_{13}(\de_{13},t)\Big]
\;\;\;;\;\;\; L_{13}(\de_{13},t)=\int
d^{3}x\lf[\nu_{1}^{\dag}(x)\nu_{3}(x)e^{-i\de_{13}}-
\nu_{3}^{\dag}(x)\nu_{1}(x)e^{i\de_{13}}\ri]\,. \eea

Six different matrices can be obtained by permuting the order of
the $G_{ij}$ (useful relations are listed in Appendix A) in
Eq.(\ref{generator}). We obtain:

\bea\non &&G_{1} \,\equiv\, G_{23}G_{13}G_{12}
\\
 &&{\cal U}_1\,=\,\begin{pmatrix}
c_{12}c_{13} & s_{12}c_{13}e^{-i\de_{12}} & s_{13}e^{-i\de_{13}} \\
-s_{12}c_{23}e^{i\de_{12}}- s_{23}s_{13}c_{12}e^{i
(\de_{13}-\de_{23}) } & c_{12}c_{23}-
s_{23}s_{13}s_{12}e^{-i(\de_{23} - \de_{13} +\de_{12})}&
s_{23}c_{13}e^{-i\de_{23}} \\
-s_{13}c_{12}c_{23}e^{i\de_{13}}+ s_{12}s_{23}e^{i (\de_{12} +
\de_{23})}&
-c_{23}s_{13}s_{12}e^{i(\de_{13}-\de_{12})}-s_{23}c_{12}e^{i\de_{23}}&
c_{23}c_{13}
\end{pmatrix}\
\eea

\bea\non &&G_{2}\,\equiv\,G_{23}G_{12}G_{13}
\\
&&{\cal U}_2\,=\,\begin{pmatrix} c_{12}c_{13} &
s_{12}e^{-i\de_{12}}
& s_{13}c_{12}e^{-i\de_{13}} \\
-s_{12}c_{13}c_{23}e^{i\de_{12}}-
s_{23}s_{13}e^{i(\de_{13}-\de_{23}) } & c_{12}c_{23} & -
s_{13}c_{23}s_{12}e^{i(\de_{12}-\de_{13})} +
s_{23}c_{13}e^{-i\de_{23}} \\
-s_{13}c_{23}e^{i\de_{13}}+
s_{12}s_{23}c_{13}e^{i(\de_{12}+\de_{23})}&
-c_{12}s_{23}e^{i\de_{23}} & c_{23}c_{13}+
s_{12}s_{13}s_{23}e^{i(\de_{12}+\de_{23}-\de_{13})}
\end{pmatrix}\
\eea

\bea\non && G_{3}\,\equiv\,G_{13}G_{23}G_{12}
\\
&&{\cal U}_3\,=\,\begin{pmatrix} c_{12}c_{13}+
s_{13}s_{23}s_{12}e^{i(\de_{12}-\de_{13}+\de_{23})} &
s_{12}c_{13}e^{-i\de_{12}}-
s_{13}s_{23}c_{12}e^{i(\de_{23}-\de_{13})}&
s_{13}c_{23}e^{-i\de_{13}} \\
-s_{12}c_{23}e^{i\de_{12}} &
c_{12}c_{23} & s_{23}e^{-i\de_{23}} \\
c_{13}s_{23}s_{12}e^{i(\de_{23}+\de_{12})}
-s_{13}c_{12}e^{i\de_{13}} & -c_{13}s_{23}c_{12}e^{i\de_{23}} -
s_{12}s_{13}e^{i(\de_{13}-\de_{12})} & c_{23}c_{13}
\end{pmatrix}\
\eea

\bea\non &&G_{4}\,\equiv\,G_{13}G_{12}G_{23}
\\
&&{\cal U}_4\,=\,\begin{pmatrix} c_{12}c_{13} &
s_{12}c_{13}c_{23}e^{-i\de_{12}} -
s_{13}s_{23}e^{i(\de_{23}-\de_{13})} &
s_{12}s_{23}c_{13}e^{-i(\de_{12}+\de_{23})}
+ s_{13}c_{23}e^{-i\de_{13}} \\
-s_{12}e^{i\de_{12}} &   c_{12}c_{23} &
s_{23}c_{12}e^{-i\de_{23}} \\
-c_{12}s_{13}e^{i\de_{13}} & -c_{13}s_{23}e^{i\de_{23}} -
s_{12}c_{23}s_{13}e^{i(\de_{13}-\de_{12})} & c_{23}c_{13} -
s_{12}s_{23}s_{13}e^{-i(\de_{12}+\de_{23}-\de_{13})}
\end{pmatrix}\
\eea

\bea\non &&G_5 \,\equiv\, G_{12}G_{13}G_{23}
\\
&&{\cal U}_5\,=\,\begin{pmatrix} c_{12}c_{13} &
s_{12}c_{23}e^{-i\de_{12}}- s_{13}c_{12}s_{23}
e^{-i(\de_{13}-\de_{23})} & s_{13}c_{12}c_{23}e^{-i\de_{13}}+
s_{12}s_{23}e^{-i(\de_{12}+\de_{23})}\\
-s_{12}c_{13}e^{i\de_{12}} & c_{12}c_{23}+
s_{12}s_{23}s_{13}e^{i(\de_{12}-\de_{13}+\de_{23})}
&s_{23}c_{12}e^{-i\de_{23}}- s_{12}c_{23}s_{13}
e^{i(\de_{12}-\de_{13})} \\
-s_{13}e^{i\de_{13}} & -c_{13}s_{23}e^{i\de_{23}} & c_{23}c_{13}
\end{pmatrix}\
\eea

\bea\ \label{matrix}\non &&G_6\,\equiv\,G_{12}G_{23}G_{13}
\\
&&{\cal U}_6\,=\,\begin{pmatrix} c_{12}c_{13}-
s_{12}s_{23}s_{13}e^{-i(\de_{12}+\de_{23} -\de_{13})} &
s_{12}c_{23} e^{-i\de_{12}} & c_{12}s_{13}e^{-i\de_{13}}+
s_{12}s_{23}c_{13}e^{-i(\de_{12}+\de_{23})} \\
-c_{12}s_{23}s_{13}e^{i(\de_{13} -\de_{23})}- s_{12}c_{13}
e^{i\de_{12}}&c_{12}c_{23} & c_{12}s_{23}c_{13}e^{-i\de_{23}}
- s_{12}s_{13}e^{i(\de_{12}-\de_{13})}\\
-c_{23}s_{13}e^{i\de_{13}} & -s_{23}e^{i\de_{23}} & c_{23}c_{13}
\end{pmatrix}\
\eea

The above matrices are generated for a particular set of initial
conditions, namely for those of Eq.(\ref{incond}). The freedom in
the choice of the initial conditions reflects into the possibility
of obtaining other unitary matrices from the above ones by
permuting rows and columns and by multiplying row or columns for a
phase factor.

We thus can easily recover all the existing parameterizations of
the CKM matrix
\cite{CKM,Chaturvedi,Fritzsch,Wolfenstein,Maiani,Chau-Keung,Anselm}:

\noi - the Maiani parameterization \cite{Chaturvedi,Maiani} is
obtained from ${\cal U}_1$ by setting $\te_{12}\rar \te$,
$\te_{13}\rar\bt$, $\te_{23}\rar \ga$, $\de_{12}\rar 0$,
$\de_{13}\rar 0$, $\de_{23}\rar -\de$;

\noi - the Chau--Keung parameterization
\cite{Chaturvedi,Chau-Keung} is recovered from ${\cal U}_1$ by
setting $\de_{12}\rar 0$ and $\de_{23}\rar 0$;

\noi - the Kobayshi--Maskawa \cite{CKM,Chaturvedi} is recovered
from ${\cal U}_5$ by setting $\te_{12}\rar \te_{2}$,
$\te_{13}\rar\te_{1}$, $\te_{23}\rar \te_{3}$, $\de_{12}\rar
-\de$, $\de_{13}\rar 0$ and $\de_{23}\rar 0$, $\te_{i}\rar
\frac{3}{2}\pi-\te_{i}$, with $i=1,2,3 $, and multiply the last
column for $(-1)$;

\noi - the Anselm parameterization \cite{Chaturvedi,Anselm} is
obtained from ${\cal U}_1$ by setting $\te_{12} \leftrightarrow
\te_{13}$, then $\de_{12}\rar 0$, $\de_{13}\rar 0$, $ \te_{12}\rar
\pi + \te_{12}$, $ \te_{13}\rar
 \pi -\te_{13}$, $ \te_{23}\rar \frac{3}{2}\pi+\te_{23}$,
exchanging second and third column and multiplying the last row
for $(-1)$.

From the above analysis it is clear that a number of new
parameterizations of the mixing matrix can be generated and that a
clear physical meaning can be attached to each of them, by
considering the order in which the generators $G_{ij}$ act and the
initial conditions used for getting that particular matrix.

\section{Currents and charges for three flavor fermion mixing}

In this Section we study the currents associated to the
Lagrangians Eq.(\ref{lag12}) and (\ref{lagemu}). To this end, let
us consider the transformations acting on the triplet of free
fields with different masses $\Psi_m$, in the line of
Ref.\cite{currents}.

${\cal L}$ is invariant under global $U(1)$ phase transformations
of the type $\Psi_m' \, =\, e^{i \al }\, \Psi_m$: as a result, we
have the conservation of the Noether charge $Q=\int d^3x \, I^0(x)
$ (with $I^\mu(x)={\bar \Psi}_m(x) \, \ga^\mu \, \Psi_m(x)$)
which is indeed the total charge of the system (i.e. the total
lepton number).

Consider then the $SU(3)$ global transformations acting on
$\Psi_m$:
\bea \label{masssu2} \Psi_m'(x) \, =\, e^{i \al_j  F_j}\, \Psi_m
(x) \, \qquad, \qquad
 j=1, 2,..., 8.
\eea
with $\al_j$ real constants, $F_{j}=\frac{1}{2}\la_{j}$ being the
generators  of $SU(3)$ and $\lambda_j$ the Gell-Mann
matrices\cite{Cheng-Li}.

The Lagrangian is not generally invariant under (\ref{masssu2})
and  we obtain, by use of the equations of motion,
\bea \non &&\de {\cal L}(x)\,= \,  i \al_j \,{\bar \Psi_m}(x)\,
[F_j,\textsf{M}_d ]\, \Psi_m(x) \, =\,  - \al_j \,\pa_\mu
J_{m,j}^\mu (x)
\\ [3mm]\label{fermacu1}
&&J^\mu_{m,j}(x)\, =\, {\bar \Psi_m}(x)\, \ga^\mu\, F_j\,
\Psi_m(x) \qquad, \qquad j=1, 2,..., 8. \eea

It is useful to list explicitly the eight currents (we suppress
spacetime dependence): \bea &&J_{m,1}^\mu \; =\; \frac{1}{2} \lf[
{\bar \nu}_1 \ga^\mu \nu_2 \, + \,{\bar \nu}_2 \ga^\mu \nu_1  \ri]
\quad \;\;,\quad J_{m,2}^\mu \; =\; -\frac{i}{2} \lf[  {\bar
\nu}_1 \ga^\mu \nu_2 \, - \,{\bar \nu}_2 \ga^\mu \nu_1  \ri]
\\
&&J_{m,3}^\mu \; =\; \frac{1}{2} \lf[  {\bar \nu}_1 \ga^\mu \nu_1
\, - \,{\bar \nu}_2 \ga^\mu \nu_2  \ri] \quad \;\;,\quad
J_{m,4}^\mu \; =\; \frac{1}{2} \lf[  {\bar \nu}_1 \ga^\mu \nu_3 \,
+ \,{\bar \nu}_3 \ga^\mu \nu_1  \ri]
\\
&&J_{m,5}^\mu \; =\; -\frac{i}{2} \lf[  {\bar \nu}_1 \ga^\mu \nu_3
\, - \,{\bar \nu}_3 \ga^\mu \nu_1  \ri] \quad ,\quad J_{m,6}^\mu
\; =\; \frac{1}{2} \lf[  {\bar \nu}_2 \ga^\mu \nu_3 \, + \,{\bar
\nu}_3 \ga^\mu \nu_2  \ri]
\\
&&J_{m,7}^\mu  \;=\; -\frac{i}{2} \lf[  {\bar \nu}_2 \ga^\mu \nu_3
\, - \,{\bar \nu}_3 \ga^\mu \nu_2  \ri] \quad ,\quad J_{m,8}^\mu
\; =\; \frac{1}{2\sqrt{3}} \lf[  {\bar \nu}_1 \ga^\mu \nu_1 \, +
\,{\bar \nu}_2 \ga^\mu \nu_2\,- 2{\bar \nu}_3 \ga^\mu \nu_3 \ri].
\eea
The related charges  $Q_{m,j}(t)\equiv \int d^3 x \,J^0_{m,j}(x)
$, satisfy the  $su(3)$  algebra $[Q_{m,j}(t), Q_{m,k}(t)]\, =\, i
\,f_{jkl}\,Q_{m,l}(t)$. Note that only two of the above charges
are time-independent, namely $Q_{m,3}$ and $Q_{m,8}$. We can thus
define the combinations:
\bea\lab{noether1} Q_{1}& \equiv &\frac{1}{3}Q \,+ \,Q_{m,3}+
\,\frac{1}{\sqrt{3}}Q_{m,8},
\\ \lab{noether2}
Q_{2}& \equiv & \frac{1}{3}Q \,-
\,Q_{m,3}+\,\frac{1}{\sqrt{3}}Q_{m,8},
\\ \lab{noether3}
Q_{3}& \equiv &\frac{1}{3}Q \,- \,\frac{2}{\sqrt{3}}Q_{m,8}, \eea
\bea\lab{charge}
 &&Q_i \, = \,\sum_{r} \int d^3 k\lf(
\al^{r\dag}_{{\bf k},i} \al^{r}_{{\bf k},i}\, -\,
\bt^{r\dag}_{-{\bf k},i}\bt^{r}_{-{\bf k},i}\ri),\,\  i=1, 2, 3 .
\eea
These are nothing but  the Noether charges associated with the
non-interacting fields $\nu_1$, $\nu_2$ and $\nu_3$: in the
absence of mixing, they are the flavor charges,  separately
conserved for each generation.

As already observed in \S II, in the case when CP is conserved
($\de=0$), the mixing generator  Eq.(\ref{generator}) is an
element of the $SU(3)$ group and can be expressed in terms of the
above charges as:
\bea \lf.G_{\bf \te}(t)\ri|_{\de=0}\, =\, e^{i 2\te_{23}
\,Q_{m,7}(t)} \, e^{i 2\te_{13} \,Q_{m,5}(t)} \, e^{i 2\te_{12}
\,Q_{m,2}(t)} \eea

Following Ref.\cite{currents}, we can now perform the $SU(3)$
transformations on the flavor triplet $\Psi_f$ and obtain another
set of currents for the flavor fields:
\bea \Psi_f'(x) \, =\, e^{i  \al_j F_j }\, \Psi_f (x) \qquad
,\qquad j \,=\, 1, 2,..., 8, \eea which leads to
\bea \non &&\de {\cal L}(x)\,= \,   i \al_j\,{\bar \Psi_f}(x)\,
[F_j, \textsf{M}]\, \Psi_f(x)\, =\, - \al_j \,\pa_\mu
J_{f,j}^{\mu}(x)\, ,
\\ [3mm] \label{fermacu2}
&&J^\mu_{f,j}(x) \,=\,  {\bar \Psi_f}(x)\, \ga^\mu\, F_j\,
\Psi_f(x)\qquad ,\qquad j \,=\, 1, 2,..., 8. \eea

Alternatively, the same currents can be obtained  by applying on
the $ J_{m,j}^{\mu}(x)$ the mixing generator
Eq.(\ref{generator}):
\bea J_{f,j}^{\mu}(x) \, = \, G_\te^{-1}(t)\,
J_{m,j}^{\mu}(x)\,G_\te(t)\qquad ,\qquad j \,=\, 1, 2,..., 8. \eea
The related charges  $Q_{f,j}(t)$ $\equiv$  $\int d^3 x
\,J^0_{f,j}(x) $ still close the $su(3)$ algebra. Due to the
off--diagonal (mixing) terms in the mass matrix $\textsf{M}$,
$Q_{f,3}(t)$ and $Q_{f,8}(t)$ are time--dependent. This implies an
exchange of charge between $\nu_e$, $\nu_\mu$ and $\nu_\tau$,
resulting in the flavor oscillations.

In accordance with Eq.(\ref{noether1}), we define the {\em flavor
charges} for mixed fields as
\bea Q_e(t) & \equiv & \frac{1}{3}Q \, + \, Q_{f,3}(t)\, +
\,\frac{1}{\sqrt{3}} Q_{f,8}(t),
\\
Q_\mu(t) & \equiv & \frac{1}{3}Q \, - \, Q_{f,3}(t)+
\,\frac{1}{\sqrt{3}} Q_{f,8}(t),
\\
Q_\tau(t) & \equiv & \frac{1}{3}Q \, -  \, \frac{2}{\sqrt{3}}
Q_{f,8}(t). \eea
with $Q_e(t) \, + \,Q_\mu(t) \,+ \,Q_\tau(t) \, = \, Q$.
These charges have a simple expression in terms of the flavor
ladder operators:
\bea\lab{flavchar} Q_\si(t) & = & \sum_{r} \int d^3 k\lf(
\al^{r\dag}_{{\bf k},\si}(t) \al^{r}_{{\bf k},\si}(t)\, -\,
\bt^{r\dag}_{-{\bf k},\si}(t)\bt^{r}_{-{\bf
k},\si}(t)\ri)\quad,\;\si= e,\mu,\tau\,, \eea
because of the connection with the Noether charges of
Eq.(\ref{charge}) via the mixing generator: $Q_\si(t) =
G^{-1}_\te(t)Q_i G_\te(t)$.

Notice also that the operator $\Delta Q_\si(t)\equiv  Q_\si(t) -
Q_i$ with $(\si,i)=(e,1) , (\mu,2) , (\tau,3)$\,, describes how
much the mixing violates the (lepton) charge conservation for a
given generation.

\vspace{0.3cm}

Let us now come back to the algebra of the currents including CP
violating phases. To this end, we consider a generalization of the
Gell-Mann matrices (we use a tilde for denoting the modified
quantities including phases):
\bea \non &&{\ti \lambda}_{1}=\begin{pmatrix}
  0 & e^{i\de_2} & 0 \\
 e^{-i\de_2} & 0 & 0 \\
  0 & 0 & 0
\end{pmatrix} \quad,\quad
{\ti \lambda}_{2}=\begin{pmatrix}
  0 & -i e^{i\de_2} & 0 \\
  i e^{-i\de_2}& 0 & 0 \\
  0 & 0 & 0
\end{pmatrix} \;,\; {\ti \lambda}_{3}=
\begin{pmatrix}
  1 & 0 & 0 \\
  0 & -1 & 0 \\
  0 & 0 & 0
\end{pmatrix}
\;, \; {\ti \lambda}_{4}=
\begin{pmatrix}
  0 & 0 & e^{-i\de_5}\\
  0 & 0 & 0 \\
  e^{i\de_5} & 0 & 0
\end{pmatrix}
\\[3mm]
&&{\ti \lambda}_{5}=\begin{pmatrix}
  0 & 0 & -ie^{-i\de_5} \\
  0 & 0 & 0 \\
  ie^{i\de_5} & 0 & 0
\end{pmatrix}\;, \;
{\ti \lambda}_{6}=\begin{pmatrix}
  0 & 0 & 0 \\
  0 & 0 & e^{i\de_7} \\
  0 & e^{-i\de_7} & 0
\end{pmatrix} \;, \; {\ti \lambda}_{7}=\begin{pmatrix}
  0 & 0 & 0 \\
  0 & 0 & -i e^{i\de_7}\\
  0 & ie^{-i\de_7} & 0
\end{pmatrix}\;, \;{\ti \lambda_{8}}=\frac{1}{\sqrt{3}}\begin{pmatrix}
  1 & 0 & 0 \\
  0 & 1 & 0 \\
  0 & 0 & -2
\end{pmatrix}.
\eea
These are normalized as the  Gell-Mann matrices: $tr(  \lambda_{j}
\lambda_{k}) =2\delta_{jk}$ . We define as usual the algebraic
generators: \bea\label{defgen} &&{\ti F}_{j}=\frac{1}{2}{\ti
\lambda}_{j}\quad,\quad j=1,..,8 \eea

The above generators Eq.(\ref{defgen}) do not close the $su(3)$
algebra, unless the condition $\de_2+\de_5+\de_7= 0$ is imposed
(cf. Eq.(\ref{com}) below), i.e. if one of the three phases is
fixed in terms of the remaining two. Such a request is clearly
incompatible with the parameterizations of the mixing matrices of
\S II and \S III (cf. e.g. the discussion after Eq.(\ref{matrix});
we have the correspondence $\{\de_2,\de_5,\de_7\}\lrar
\{\de_{12},\de_{13},\de_{23}\}$). For the matrix (\ref{fermix}),
as already observed, it implies $\de=0$.

The ${\ti F}_{j}$ satisfy a deformed $su(3)$ algebra with
deformed commutation relations given by:
\bea \label{com}\non &&[{\ti F}_2,{\ti F}_5]\, =\, \frac{i}{2} \,
{\ti F}_7 \, e^{-i \,\De\, ({\ti F}_3-\sqrt{3}{\ti F}_8)}
 \quad;\quad [{\ti F}_2,{\ti F}_7]\, =\, -\frac{i}{2}
\, {\ti F}_5 \, e^{-i \,\De\,({\ti F}_3+\sqrt{3}{\ti F}_8) }
 \quad ; \quad [{\ti F}_5,{\ti F}_7]\, =\, \frac{i}{2} \, {\ti F}_2\,
 e^{ 2i\,\De\, {\ti F}_3}
\\ \non
&&[{\ti F}_1,{\ti F}_4]\, =\, \frac{i}{2} \, {\ti F}_7 \, e^{-i
\,\De\,({\ti F}_3-\sqrt{3}{\ti F}_8) } \quad;\quad [{\ti F}_1,{\ti
F}_7]\, =-\, \frac{i}{2} \, {\ti F}_4 \, e^{- i \,\De\,({\ti
F}_3+\sqrt{3}{\ti F}_8) } \quad ; \quad [{\ti F}_4,{\ti F}_7]\,
=\, \frac{i}{2} \, {\ti F}_1\, e^{2 i \,\De\,{\ti F}_3}
\\ \non
&&[{\ti F}_1,{\ti F}_5]\, =\,- \frac{i}{2} \, {\ti F}_6 \, e^{-i
\,\De\,({\ti F}_3-\sqrt{3}{\ti F}_8)} \quad ; \quad [{\ti
F}_1,{\ti F}_6]\, =\, \frac{i}{2} \, {\ti F}_5 \, e^{-i \,\De\,
({\ti F}_3+\sqrt{3}{\ti F}_8)} \quad ; \quad [{\ti F}_5,{\ti
F}_6]\, =\,- \frac{i}{2} \, {\ti F}_1 \, e^{2i \,\De\,{\ti F}_3}
\\
&&[{\ti F}_2,{\ti F}_4]\, =\, \frac{i}{2} \, {\ti F}_6\, e^{-i
\,\De\, ({\ti F}_3-\sqrt{3}{\ti F}_8)} \quad ; \quad [{\ti
F}_2,{\ti F}_6]\, =\, - \frac{i}{2} \, {\ti F}_4\, e^{-i
\,\De\,({\ti F}_3+\sqrt{3}{\ti F}_8)} \quad ; \quad [{\ti
F}_4,{\ti F}_6]\, =\, \frac{i}{2} \, {\ti F}_2\, e^{2 i
\,\De\,{\ti F}_3} \eea
where  $\De \equiv \de_2+\de_5 +\de_7$. The other commutators are
the usual $su(3)$ ones. For $\De=0$, the $su(3)$ algebra is
recovered.

It is useful to look at the deformed algebra in terms of the
raising and lowering operators, defined as\cite{Cheng-Li}:
\bea && {\ti T}_\pm \equiv {\ti F}_1 \pm i {\ti F}_2 \quad, \quad
{\ti U}_\pm \equiv {\ti F}_6 \pm i {\ti F}_7 \quad, \quad {\ti
V}_\pm \equiv {\ti F}_4 \pm i {\ti F}_5 \eea
We also define:
\bea && {\ti T}_3 \equiv  {\ti F}_3 \quad, \quad {\ti U}_3 \equiv
\frac{1}{2}\lf(\sqrt{3} {\ti F}_8 - {\ti F}_3 \ri) \quad, \quad
{\ti V}_3 \equiv \frac{1}{2}\lf(\sqrt{3} {\ti F}_8 + {\ti F}_3
\ri) \eea
Then the only deformed commutators are the following ones: \bea &&
[{\ti T}_+,{\ti V}_-] \,=\, - {\ti U}_- \,e^{2 i \De {\ti U}_3}
\quad, \quad [{\ti T}_+,{\ti U}_+] \,=\,  {\ti V}_+ \,e^{-2 i \De
{\ti V}_3} \quad, \quad [{\ti U}_+,{\ti V}_-] \,=\,  {\ti T}_-
\,e^{2 i \De {\ti T}_3} \eea

\section{Neutrino Oscillations}

 The oscillation formulas are
obtained by taking expectation values of the above charges on the
(flavor) neutrino state. Consider for example an initial electron
neutrino state defined as $|\nu_e\ran \equiv \al_{{\bf k},e}^{r
\dag}(0) |0\ran_{f}$ (for a discussion on the  correct definition
of flavor states see Refs.\cite{BHV99,remarks,bosonmix}). Working
in the Heisenberg picture, we obtain
\bea \label{charge1} {\cal Q}^\rho_{{\bf k},\si}(t) &\equiv&
\langle \nu_\rho|Q_\si(t)| \nu_\rho\ran - \, {}_f\lan 0 |Q_\si(t)|
0\ran_f \;=\, \lf|\lf \{\al^{r}_{{\bf k},\si}(t), \al^{r
\dag}_{{\bf k},\rho}(0) \ri\}\ri|^{2} \;+ \;\lf|\lf\{\bt_{{-\bf
k},\si}^{r \dag}(t), \al^{r \dag}_{{\bf k},\rho}(0) \ri\}\ri|^{2}
\,,
\\ [2mm] \label{charge2}
{\cal Q}^{\bar \rho}_{{\bf k},\si}(t) &\equiv& \langle {\bar
\nu}_\rho|Q_\si(t)| {\bar \nu}_\rho\ran - \, {}_f\lan 0 |Q_\si(t)|
0\ran_f \;=\, - \lf|\lf \{\bt^{r}_{{\bf k},\si}(t), \bt^{r
\dag}_{{\bf k},\rho}(0) \ri\}\ri|^{2} \;- \;\lf|\lf\{\al_{{-\bf
k},\si}^{r \dag}(t), \bt^{r \dag}_{{\bf k},\rho}(0) \ri\}\ri|^{2}
\,, \eea
where $|0\ran_{f}\equiv |0(0)\ran_{f}$. Overall charge
conservation is obviously ensured at any time: ${\cal Q}_{{\bf
k},e}(t) + {\cal Q}_{{\bf k},\mu}(t)+{\cal Q}_{{\bf k},\tau}(t) \;
= \; 1$. We remark that the expectation value of $Q_\si$ cannot be
taken on vectors of the Fock space built on $|0\ran_{m}$, as shown
in Refs.\cite{BHV99,remarks,bosonmix}. Also we observe that
${}_f\lan 0 |Q_\si(t)| 0\ran_f  \neq 0$, in contrast with the two
flavor case\cite{remarks,comment}. We introduce the following
notation:
\bea\non \De_{ij}^{\bf k}\equiv \frac{\om_{k,j} - \om_{k,i}}{2}
\quad , \quad \Om_{ij}^{\bf k}\equiv \frac{\om_{k,i} +
\om_{k,j}}{2} \eea
Then the  oscillation (in time) formulas for the flavor charges,
on an initial electron neutrino state, follow as:
\bea \non {\cal Q}^e_{{\bf k},e}(t) \, &=& \,1 \,-\, \sin^{2}( 2
\te_{12})\cos^{4}\te_{13} \, \Big[|U_{12}^{\bf k}|^2\, \sin^{2}
\lf(  \De_{12}^{\bf k}  t \ri)
 +\,|V_{12}^{\bf k}|^2 \, \sin^{2}
\lf(\Om_{12}^{\bf k}t \ri)\Big] \
\\ \non && -\, \sin^{2}(2 \te_{13})\cos^{2}\te_{12}
\, \Big[|U_{13}^{\bf k}|^2\,
 \sin^{2} \lf( \De_{13}^{\bf k}  t \ri)
 +\,|V_{13}^{\bf k}|^2\, \sin^{2} \lf( \Om_{13}^{\bf k} t \ri)\Big] \
\\ &&
-\, \sin^{2}(2 \te_{13})\sin^{2}\te_{12} \, \Big[|U_{23}^{\bf
k}|^2 \, \sin^{2} \lf( \De_{23}^{\bf k} t \ri) +\,|V_{23}^{\bf
k}|^2\, \sin^{2} \lf( \Om_{23}^{\bf k} t \ri)\Big]\,, \eea
\bea\non {\cal Q}^e_{{\bf k},\mu}(t)
 &=& 2 J_{\CP}
 \Big[|U_{12}^{\bf k}|^2\, \sin(2\De_{12}^{\bf k}t)
- |V_{12}^{\bf k}|^2\, \sin(2\Om_{12}^{\bf k} t) + (|U_{12}^{\bf
k}|^2 - |V_{13}^{\bf k}|^2 ) \sin(2\De_{23}^{\bf k}t)
\\ \non
&+& (|V_{12}^{\bf k}|^2 - |V_{13}^{\bf k}|^2 ) \sin(2\Om_{23}^{\bf
k}t)
  - |U_{13}^{\bf k}|^2\,
\sin(2\De_{13}^{\bf k}t)+ |V_{13}^{\bf k}|^2\, \sin(2\Om_{13}^{\bf
k}t)\Big]
\\ \non
&+&\, \cos^{2}\te_{13} \sin\te_{13}
\Big[\cos\de\sin(2\te_{12})\sin(2\te_{23}) + 4
\cos^2\te_{12}\sin\te_{13}\sin^2\te_{23}\Big]\Big[|U_{13}^{\bf
k}|^2\sin^{2} \lf(\De_{13}^{\bf k} t \ri) + |V_{13}^{\bf k}|^2\
\sin^{2} \lf( \Om_{13}^{\bf k} t \ri)\Big]
\\ \non
& -& \cos^{2}\te_{13}\sin\te_{13}
 \Big[\cos\de\sin(2\te_{12})\sin(2\te_{23}) -
4 \sin^2\te_{12}\sin\te_{13}\sin^2\te_{23}\Big] \Big[|U_{23}^{\bf
k}|^2\ \sin^{2} \lf( \De_{23}^{\bf k} t \ri)
 + |V_{23}^{\bf k}|^2\
\sin^{2} \lf( \Om_{23}^{\bf k} t \ri)\Big]
\\ \non
& +&\cos^{2}\te_{13} \sin(2\te_{12}) \Big[ (\cos^2\te_{23} -
\sin^2\te_{23}\sin^2\te_{13})\sin(2\te_{12})
+\cos\de\cos(2\te_{12})\sin\te_{13}\sin(2\te_{23})\Big] \times
\\
&\times& \Big[|U_{12}^{\bf k}|^2\ \sin^{2} \lf(\De_{12}^{\bf k} t
\ri) + |V_{12}^{\bf k}|^2\ \sin^{2} \lf( \Om_{12}^{\bf k} t
\ri)\Big]\,, \eea
\bea\non {\cal Q}^e_{{\bf k},\tau}(t)&=& - 2 J_{\CP}
 \Big[|U_{12}^{\bf k}|^2\ \sin(2\De_{12}^{\bf k}t)
- |V_{12}^{\bf k}|^2\, \sin(2\Om_{12}^{\bf k} t) + (|U_{12}^{\bf
k}|^2\, - |V_{13}^{\bf k}|^2 ) \sin(2\De_{23}^{\bf k}t)
\\ \non
&+& (|V_{12}^{\bf k}|^2\, - |V_{13}^{\bf k}|^2 )
\sin(2\Om_{23}^{\bf k}t)
 \, - |U_{13}^{\bf k}|^2\,
\sin(2\De_{13}^{\bf k}t)+ |V_{13}^{\bf k}|^2\, \sin(2\Om_{13}^{\bf
k}t)\Big]
\\ \non
&-& \cos^{2}\te_{13}\sin\te_{13}
\Big[\cos\de\sin(2\te_{12})\sin(2\te_{23}) -4
\cos^2\te_{12}\sin\te_{13}\cos^2\te_{23}\Big] \Big[|U_{13}^{\bf
k}|^2\, \sin^{2} \lf( \De_{13}^{\bf k} t \ri) + |V_{13}^{\bf
k}|^2\, \sin^{2}\lf(\Om_{13}^{\bf k} t\ri)\Big]
\\ \non
& +&\cos^{2}\te_{13}\sin\te_{13}
 \Big[\cos\de\sin(2\te_{12})\sin(2\te_{23}) +
  4 \sin^2\te_{12}\sin\te_{13}\cos^2\te_{23}\Big]\Big[|U_{23}^{\bf k}|^2 \,
\sin^{2} \lf(\De_{23}^{\bf k} t \ri) +
 |V_{23}^{\bf k}|^2\,\sin^{2} \lf(\Om_{23}^{\bf k}  t \ri)\Big]
\\ \non
&+& \cos^{2}\te_{13} \sin(2\te_{12}) \Big[ (\sin^2\te_{23} -
\sin^2\te_{13}\cos^2\te_{23})\sin(2\te_{12})
-\cos\de\cos(2\te_{12})\sin\te_{13}\sin(2\te_{23})\Big] \times
\\
&\times & \Big[|U_{12}^{\bf k}|^2\, \sin^{2} \lf( \De_{12}^{\bf k}
t \ri) + |V_{12}^{\bf k}|^2\, \sin^{2} \lf(\Om_{12}^{\bf k} t
\ri)\Big]\,, \eea
where we used the relations Eqs.(\ref{ident1})-(\ref{ident4}).
We also introduced the Jarlskog factor $J_{\CP}$ defined
as\cite{Jarlskog}
\bea J_{\CP} \equiv Im (u_{i\alpha}u_{j \beta}u^{*}_{i
\beta}u^{*}_{j\alpha}), \eea
where the $u_{ij}$ are the elements of mixing matrix ${\cal U}$
and $i \neq j, \,\ \alpha \neq \beta$. In the parameterization
Eq.(\ref{fermix}), $J_{\CP}$ is given by
 \bea
J_{\CP} = \frac{1}{8}\, \sin \de\, \sin(2\te_{12})
\sin(2\te_{13})\cos\te_{13} \sin(2\te_{23})  . \eea
Evidently, $J_{\CP}$ vanishes if $ \theta_{ij}=0,\pi/2$ and/or
$\de=0,\pi$: all CP--violating effects are proportional to it.

The above oscillation formulas are exact. The differences with
respect to the usual formulas for neutrino oscillations are in the
energy dependence of the amplitudes and in the additional
oscillating terms. For $|{\bf k}|\gg\sqrt{m_1m_2}$, we have
$|U_{ij}^{{\bf k}}|^{2}\rar 1$ and $|V_{ij}^{{\bf k}}|^{2}\rar 0$
and the traditional (Pontecorvo) oscillation formulas are
approximately recovered. Indeed, for sufficiently small time
arguments, a correction to the Pontecorvo formula is present even
in the relativistic limit.

In Appendix B the oscillation formulas for the flavor charges on
an initial electron anti-neutrino state are given.

We plot in Figs.(2) and(4) the QFT oscillation formulas ${\cal
Q}^e_{{\bf k},e}(t)$ and ${\cal Q}^e_{{\bf k},\mu}(t)$ in function
of time, and in Figs. (3) and (5)  the corresponding Pontecorvo
oscillation formulas $P^{\bf k}_{e\rightarrow e}(t)$ and $P^{\bf
k}_{e\rightarrow \mu}(t)$. The time scale is in $T_{12}$ units,
where $T_{12}= \pi/\De_{12}^{\bf k}$ is, for the values of
parameters of Tab.(1), the largest oscillation period.

\vspace{1cm} \centerline{\epsfysize=3.0truein\epsfbox{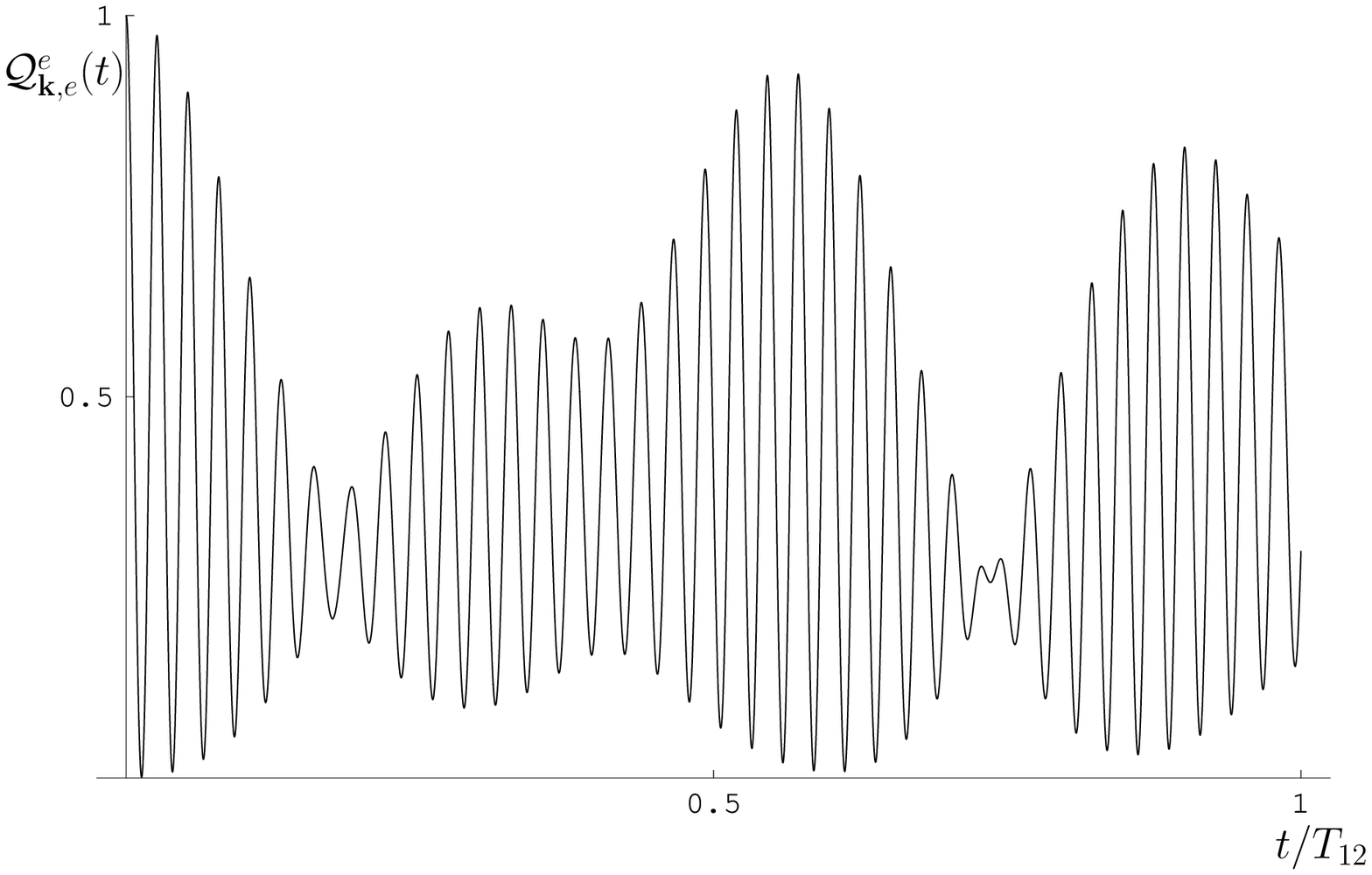}}
\vspace{.2cm} \centerline{\small Figure 2:  Plot of QFT
oscillation formula: ${\cal Q}^e_{{\bf k},e}(t)$ in function of
time for $k =55$ and parameters as in Tab.(1). }

\vspace{0.5cm}

\centerline{\epsfysize=3.0truein\epsfbox{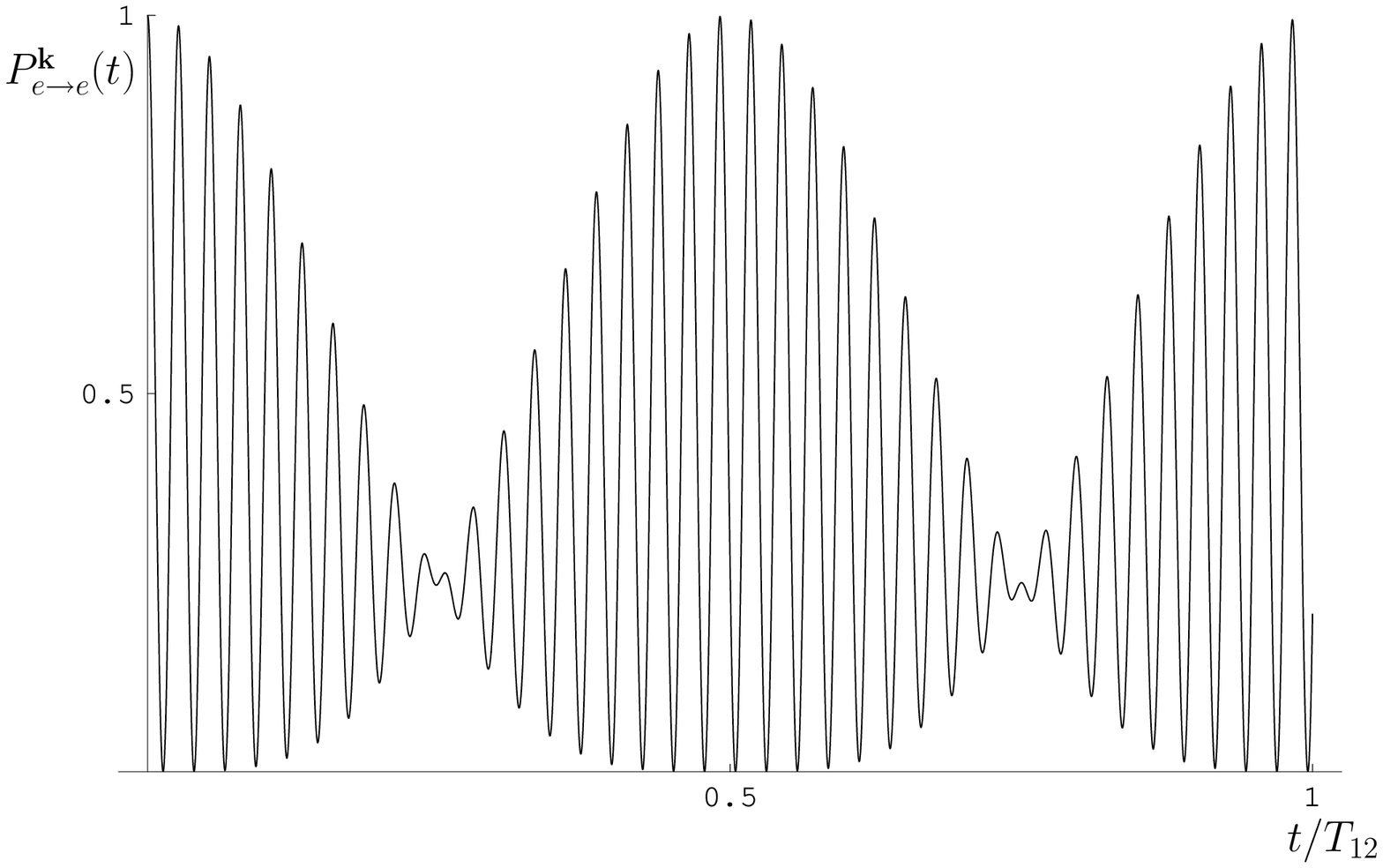}} \vspace{.2cm}
\centerline{\small Figure 3:  Plot of QM oscillation formula:
$P^{\bf k}_{e\rightarrow e}(t)$ in function of time for $k =55$
and parameters as in Tab.(1). }

\vspace{0.5cm}

\centerline{\epsfysize=3.0truein\epsfbox{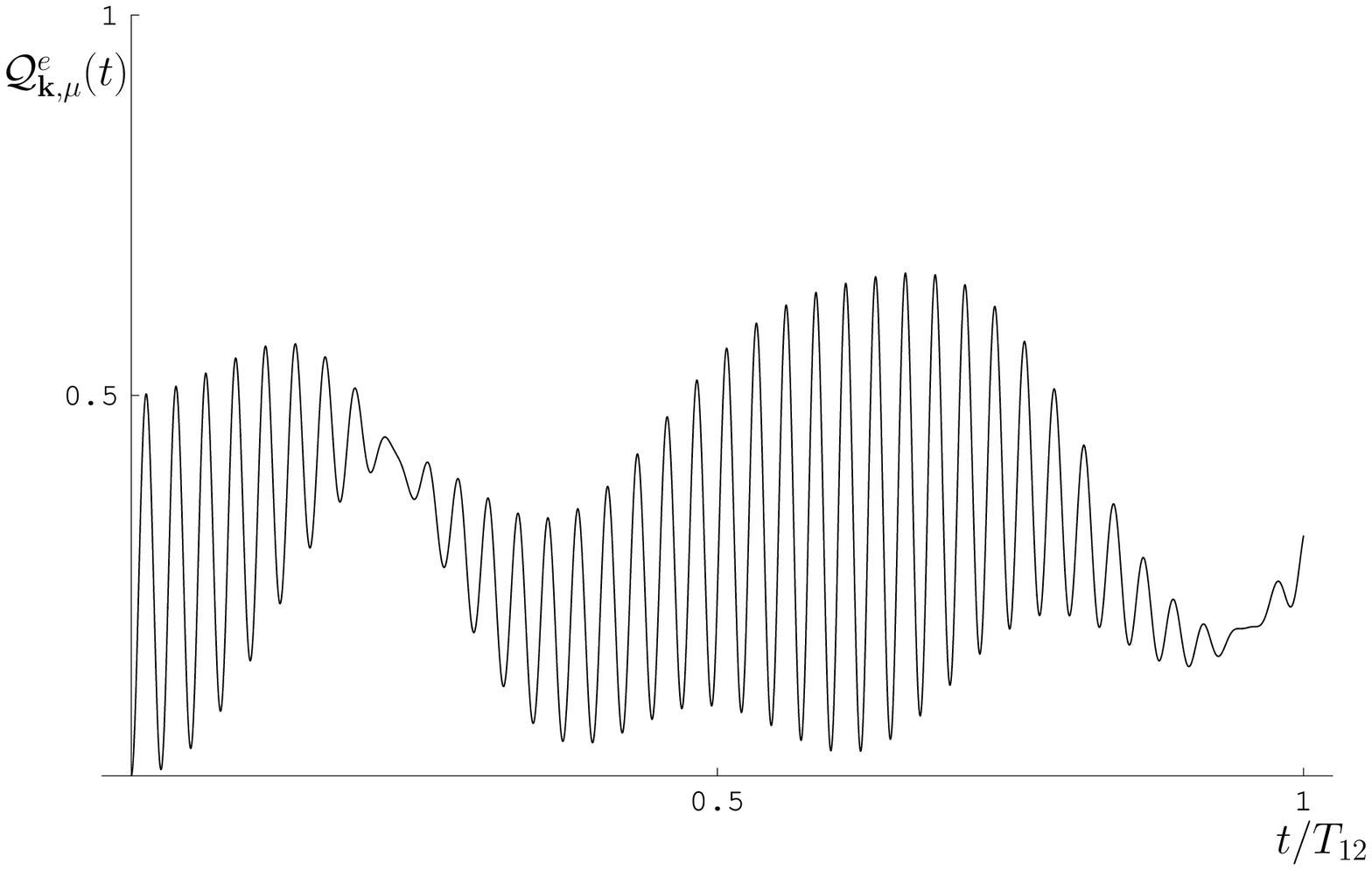}} \vspace{.2cm}
\centerline{\small Figure 4:  Plot of QFT oscillation formula:
${\cal Q}^e_{{\bf k},\mu}(t)$ in function of time for $k =55$ and
parameters as in Tab(1). }

\vspace{0.5cm}

\centerline{\epsfysize=3.0truein\epsfbox{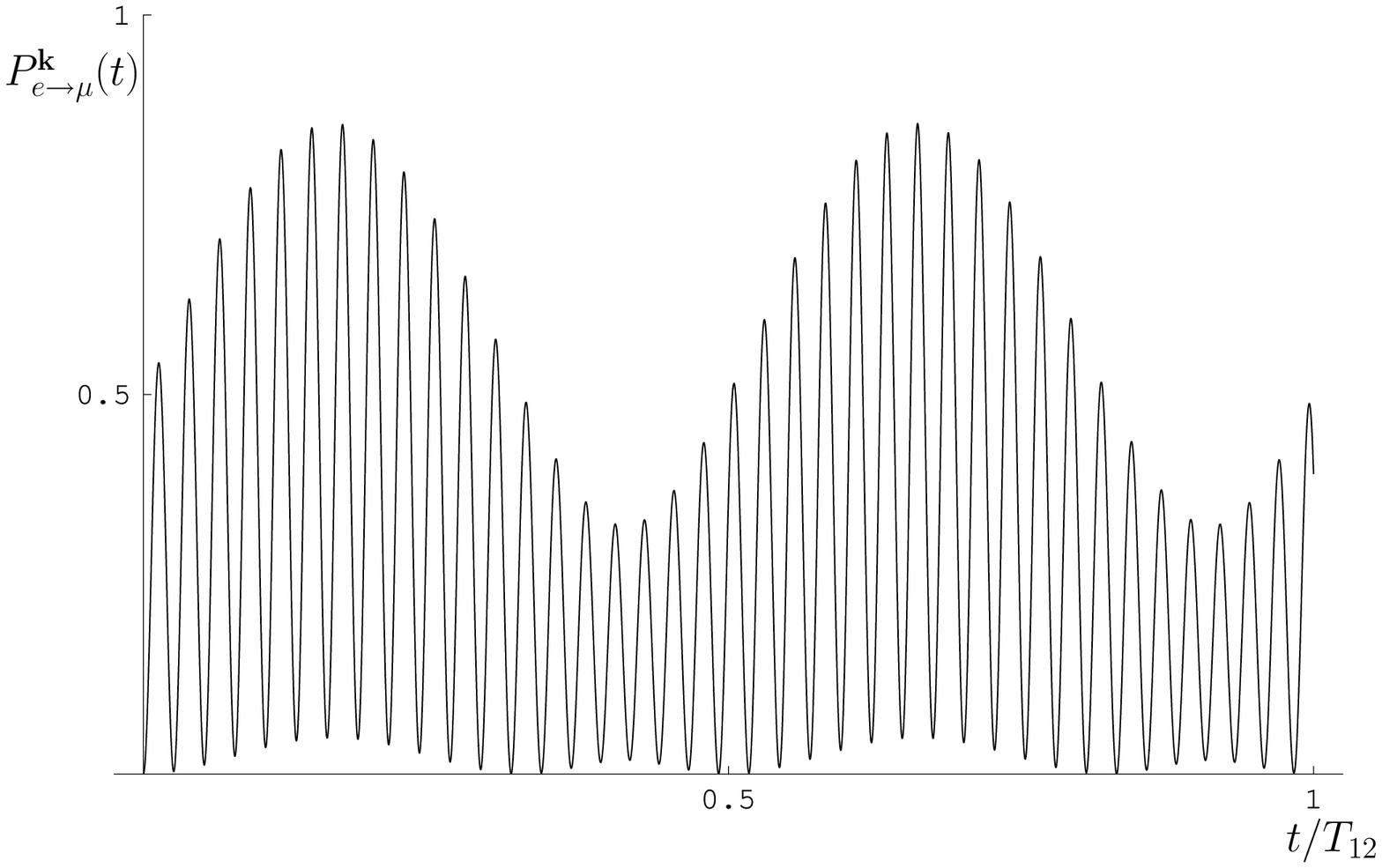}} \vspace{.2cm}
\centerline{\small Figure 5:  Plot of QM oscillation formula:
$P^{\bf k}_{e\rightarrow \mu}(t)$ in function of time for $k =55$
and parameters as in Tab.(1). }

\vspace{0.5cm}

\section{CP and T violations in neutrino oscillations}

In this Section we consider the oscillation induced CP and T
violation in the context of the present QFT framework. Let us
first briefly  recall the situation in QM\footnote{We use here the
"hat"  for QM quantities. For notational simplicity, we also
suppress momentum indices where unnecessary.}: there, the CP
asymmetry between the probabilities of two conjugate neutrino
transitions, due to CPT invariance and unitarity of the mixing
matrix, is given as \cite{Fritzsch}
\bea {\hat \Delta}_{\CP}^{\rho\si}(t)\equiv P_ {\nu_{\sigma}\rar
\nu_{\rho}}(t) - P_ {\overline{\nu}_{\sigma}\rar
\overline{\nu}_{\rho}}(t),
 \eea
 where
$ \sigma, \rho = e, \mu, \tau.$ The T violating asymmetry can be
obtained in similar way as \cite{Fritzsch}
\bea\lab{DeT} {\hat \Delta}_{\T}^{\rho\si}(t)\,\equiv\, P_
{\nu_{\sigma}\rar \nu_{\rho}}(t) - P_ {\nu_{\rho}\rar
\nu_{\sigma}}(t)\, = \,P_ {\nu_{\sigma}\rar \nu_{\rho}}(t) - P_
{\nu_{\sigma}\rar \nu_{\rho}}(-t)\,. \eea
The relationship $ {\hat \Delta}_{\CP}^{\rho\si}(t)= {\hat
\Delta}_{\T}^{\rho\si}(t)$ is a consequence of CPT invariance.

The corresponding quantities in QFT have to be defined in the
framework of the previous Section, i.e. as expectation values of
the flavor charges on states belonging to the flavor Hilbert
space. We thus have for the CP violation:
\bea\lab{DECP} \De^{\rho\si}_\CP(t) & \equiv &
 {\cal Q}^\rho_{{\bf k},\si}(t)  \, +\,
{\cal Q}^{\bar \rho}_{{\bf k},\si}(t)
\\
&=& \lf|\lf \{\al^{r}_{{\bf k},\si}(t), \al^{r \dag}_{{\bf
k},\rho}(0) \ri\}\ri|^{2} \,+ \,\lf|\lf\{\bt_{{-\bf k},\si}^{r
\dag}(t), \al^{r \dag}_{{\bf k},\rho}(0) \ri\}\ri|^{2} \,-\,
\lf|\lf \{\al^{r\dag}_{-{\bf k},\si}(t), \bt^{r \dag}_{{\bf
k},\rho}(0) \ri\}\ri|^{2} \,- \,\lf|\lf\{\bt_{{\bf k},\si}^{r}(t),
\bt^{r \dag}_{{\bf k},\rho}(0) \ri\}\ri|^{2}\,. \eea

We have
\bea &&\sum_\si \De^{\rho\si}_\CP \, = \,0 \qquad, \quad
\rho,\si=e,\mu,\tau, \eea
which follows from the fact that $\sum_\si Q_\si(t) =Q$  and
$\langle \nu_\rho|Q| \nu_\rho\ran\,=\,1$ and $\langle {\bar
\nu}_\rho|Q| {\bar \nu}_\rho\ran \,=\,-1$.

\vspace{0.2cm}

We can calculate the CP asymmetry Eq.(\ref{DECP}) for a specific
case, namely for the transition $\nu_e \longrightarrow \nu_\mu $.
We obtain
\bea\non \De^{e\mu}_\CP(t) &=&4 J_{\CP}
 \Big[|U_{12}^{\bf k}|^2\, \sin(2 \De_{12}^{\bf k}t)
- |V_{12}^{\bf k}|^2\, \sin(2 \Om_{12}^{\bf k} t) + (|U_{12}^{\bf
k}|^2 - |V_{13}^{\bf k}|^2 ) \sin(2 \De_{23}^{\bf k}t)
\\\lab{Demt}
&& + (|V_{12}^{\bf k}|^2 - |V_{13}^{\bf k}|^2 ) \sin(2
\Om_{23}^{\bf k}t)
  - |U_{13}^{\bf k}|^2\,
\sin(2 \De_{13}^{\bf k}t)+ |V_{13}^{\bf k}|^2\, \sin(2
\Om_{13}^{\bf k}t)\Big]\,, \eea
and $ \De^{e\tau}_\CP(t)= - \De^{e\mu}_\CP (t)$. As already
observed for oscillation formulas, high-frequency oscillating
terms and Bogoliubov coefficients in the oscillation amplitudes
appear in Eq.(\ref{Demt}) as a QFT correction to the QM formula.

The definition of the QFT analogue of the T-violating quantity
Eq.(\ref{DeT}) is more delicate. Indeed,
 defining  $\De_\T$ as
$ \De^{e\mu}_\T \equiv {\cal Q}^e_{\mu}(t)  \, -\, {\cal
Q}^\mu_{e}(t) $ does not seem to work, since we obtain
$\De^{e\mu}_\T \, - \,\De^{e\mu}_\CP \, \neq \, 0 $ in contrast
with CPT conservation.

A more consistent definition of the time-reversal violation in QFT
is then:
\bea
 \De^{\rho\si}_\T (t) \, \equiv
\, {\cal Q}^\rho_{{\bf k},\si}(t)  \, -\, {\cal Q}^\rho_{{\bf
k},\si}(-t)\qquad, \quad \rho,\si=e,\mu,\tau\,. \eea
With such definition, the equality $\De^{\rho\si}_\T(t)
\,=\,\De^{\rho\si}_\CP(t)$ follows from ${\cal Q}^\rho_{{\bf
k},\si}(-t)  \, =\, - {\cal Q}^{\bar \rho}_{{\bf k},\si}(t)$.

We plot in  Fig.(6)  the CP asymmetry Eq.(\ref{Demt}) for sample
values of the parameters as in Tab.(1). In Fig.(7) the
corresponding standard QM quantity is plotted for the same values
of parameters.

\vspace{1cm}

\centerline{\epsfysize=3.0truein\epsfbox{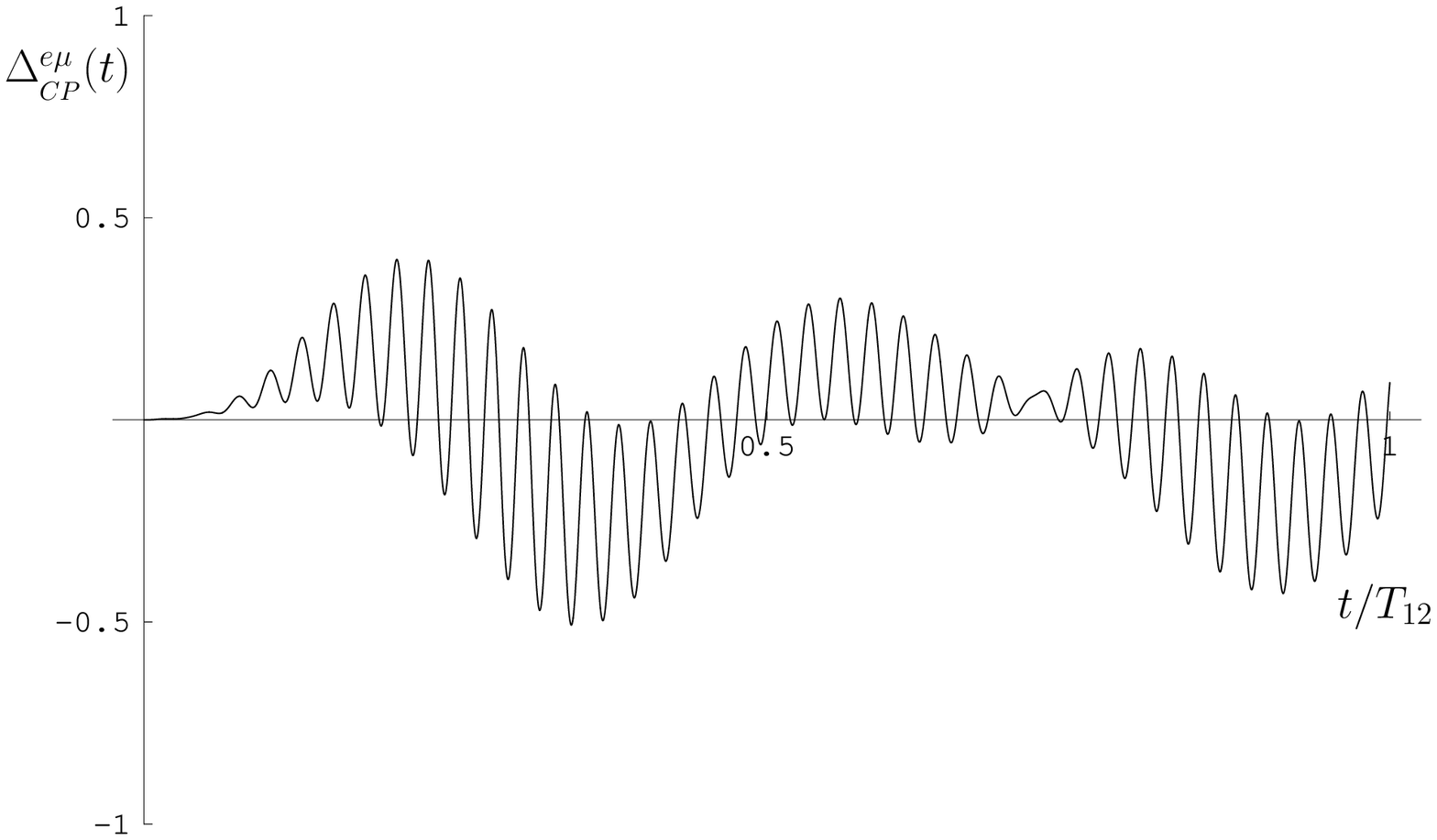}} \vspace{.2cm}
\centerline{\small Figure 6:  Plot of the QFT CP asymmetry
$\Delta^{e\mu}_{\CP}(t)$, in function of time for $k =55$ and
parameters as in Tab.(1). }

\vspace{0.5cm}

\centerline{\epsfysize=3.0truein\epsfbox{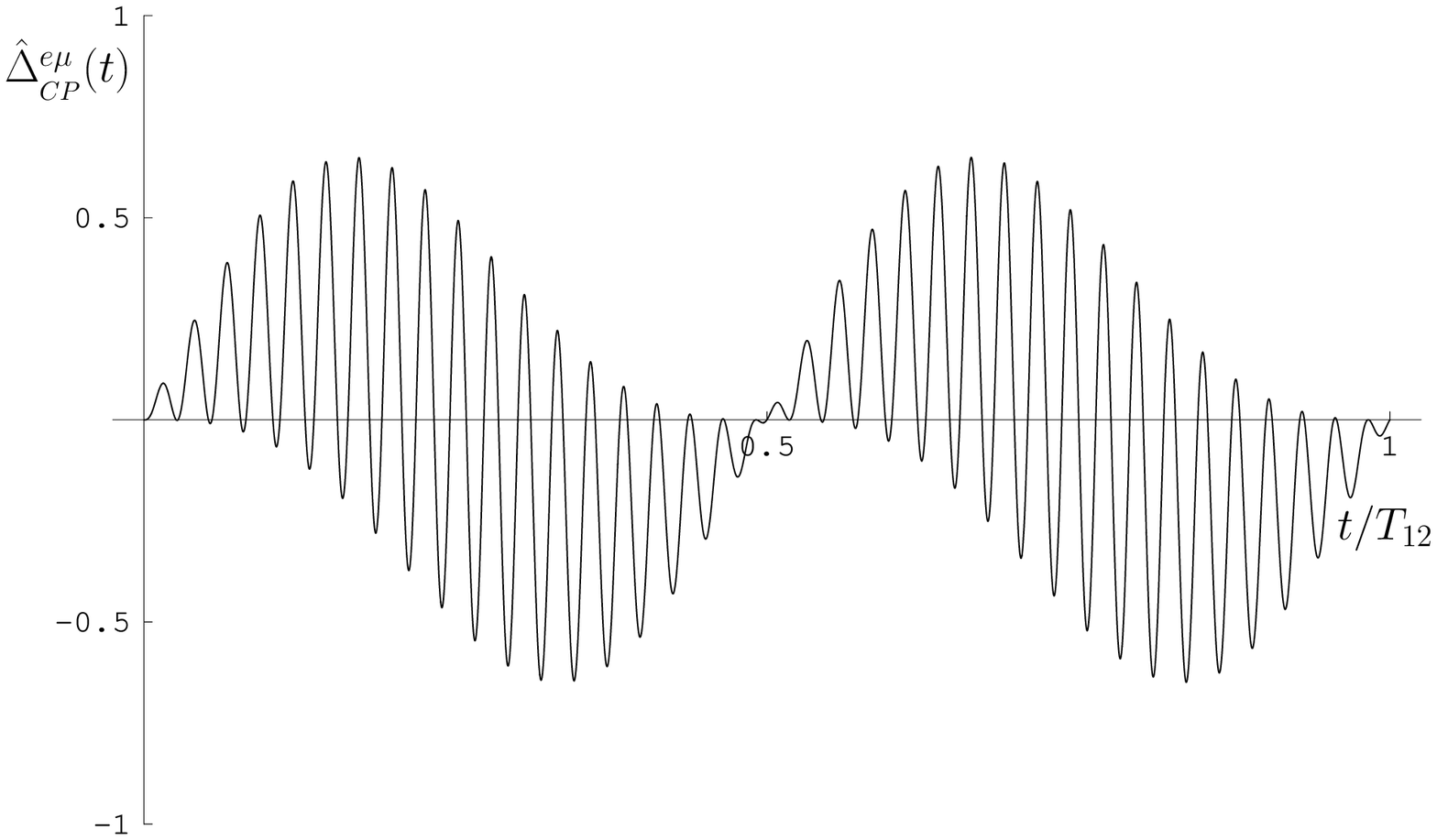}} \vspace{.2cm}
\centerline{\small Figure 7:  Plot of the QM CP asymmetry ${\hat
\Delta}^{e\mu}_{\CP}(t)$, in function of time for $k =55$
 and parameters as in Tab.(1). }

\vspace{0.5cm}

\section{Conclusions}

In this paper, we have discussed the mixing of (Dirac) fermionic
fields in Quantum Field Theory for the case of three flavors with
CP violation. We constructed the flavor Hilbert space and studied
the currents and charges for mixed fields (neutrinos). The
algebraic structure associated with the mixing for the case of
three generation  turned out to be that of a deformed $su(3)$
algebra, when a CP violating phase is present.

We have then derived all the known parameterization of the
three-flavor mixing matrix and a number of new ones. We have shown
that these parameterizations actually reflect the group
theoretical structure of the generator of the mixing
transformations.

By use of the flavor Hilbert space, we have calculated the exact
QFT oscillation formulas, a generalization of the usual QM
Pontecorvo formulas. The comparison between the exact oscillation
formulas and the usual ones has been explicitly exhibited for
sample values of the neutrino masses and mixings. CP and T
violation induced by neutrino oscillations have also been
discussed.

We remark that the corrections introduced by the present formalism
to the usual Pontecorvo formulas are in principle experimentally
testable. The fact that these corrections may be quantitatively
below the experimental accuracy reachable at the present state of
the art in the detection of the neutrino oscillations, does not
justify neglecting them in the analysis of the particle mixing and
oscillation mechanism. The exact oscillation formulas here derived
are the result of a mathematically consistent analysis which
cannot be ignored in a correct treatment of the field mixing
phenomenon. As we have seen above, our formalism accounts for all
the known parameterizations of the mixing matrix and explains
their origin and their reciprocal relations, thus unifying the
phenomenological proposals scattered in the literature where such
parameterizations have been presented. Moreover, our formalism
clearly points to the truly non-perturbative character of the
particle mixing phenomenon. A lot of Physics must be there waiting
to be discovered.

\section*{Acknowledgements}

The present research has been carried out in the framework of the
ESF network COSLAB and we acknowledge partial financial support
also by INFN, INFM and EPSRC.

\section*{Appendix A: Anti-neutrino oscillation formulas}

If we consider an initial electron anti-neutrino state defined as
$|\overline{\nu}_e\ran \equiv \bt_{{\bf k},e}^{r\dag}(0)
|0\ran_{f}$, we obtain the anti-neutrino oscillation formulas as
\bea  {\cal Q}^{\bar e}_{{\bf k},e}(t)
 \,= - {\cal Q}^{ e}_{{\bf k},e}(t)\,,
\eea

\bea\non {\cal Q}^{\bar e}_{{\bf k},\mu}(t) &=&2 J_{\CP}
 \Big[|U_{12}^{\bf k}|^2\, \sin(2\De_{12}^{\bf k}t)
- |V_{12}^{\bf k}|^2\, \sin(2\Om_{12}^{\bf k} t) + (|U_{12}^{\bf
k}|^2 - |V_{13}^{\bf k}|^2 ) \sin(2\De_{23}^{\bf k}t)
\\ \non
&+& (|V_{12}^{\bf k}|^2 - |V_{13}^{\bf k}|^2 ) \sin(2\Om_{23}^{\bf
k}t)
  - |U_{13}^{\bf k}|^2\,
\sin(2\De_{13}^{\bf k}t)+ |V_{13}^{\bf k}|^2\, \sin(2\Om_{13}^{\bf
k}t)\Big]
\\ \non
&-&\, \cos^{2}\te_{13} \sin\te_{13}
\Big[\cos\de\sin(2\te_{12})\sin(2\te_{23}) + 4
\cos^2\te_{12}\sin\te_{13}\sin^2\te_{23}\Big]\Big[|U_{13}^{\bf
k}|^2\sin^{2} \lf(\De_{13}^{\bf k} t \ri) + |V_{13}^{\bf k}|^2\
\sin^{2} \lf( \Om_{13}^{\bf k} t \ri)\Big]
\\ \non
& +& \cos^{2}\te_{13}\sin\te_{13}
 \Big[\cos\de\sin(2\te_{12})\sin(2\te_{23}) -
4 \sin^2\te_{12}\sin\te_{13}\sin^2\te_{23}\Big] \Big[|U_{23}^{\bf
k}|^2\ \sin^{2} \lf( \De_{23}^{\bf k} t \ri)
 + |V_{23}^{\bf k}|^2\
\sin^{2} \lf( \Om_{23}^{\bf k} t \ri)\Big]
\\ \non
& -&\cos^{2}\te_{13} \sin(2\te_{12}) \Big[ (\cos^2\te_{23} -
\sin^2\te_{23}\sin^2\te_{13})\sin(2\te_{12})
+\cos\de\cos(2\te_{12})\sin\te_{13}\sin(2\te_{23})\Big] \times
\\
&\times& \Big[|U_{12}^{\bf k}|^2\ \sin^{2} \lf(\De_{12}^{\bf k} t
\ri) + |V_{12}^{\bf k}|^2\ \sin^{2} \lf( \Om_{12}^{\bf k} t
\ri)\Big]\, , \eea

\bea \non {\cal Q}^{\bar e}_{{\bf k},\tau}(t) &=& - 2 J_{\CP}
 \Big[|U_{12}^{\bf k}|^2\ \sin(2\De_{12}^{\bf k}t)
- |V_{12}^{\bf k}|^2\, \sin(2\Om_{12}^{\bf k} t) + (|U_{12}^{\bf
k}|^2\, - |V_{13}^{\bf k}|^2 ) \sin(2\De_{23}^{\bf k}t)
\\ \non
&+& (|V_{12}^{\bf k}|^2\, - |V_{13}^{\bf k}|^2 )
\sin(2\Om_{23}^{\bf k}t)
 \, - |U_{13}^{\bf k}|^2\,
\sin(2\De_{13}^{\bf k}t)+ |V_{13}^{\bf k}|^2\, \sin(2\Om_{13}^{\bf
k}t)\Big]
\\ \non
&+& \cos^{2}\te_{13} \sin\te_{13}
\Big[\cos\de\sin(2\te_{12})\sin(2\te_{23}) -4
\cos^2\te_{12}\sin\te_{13}\cos^2\te_{23} \Big] \Big[|U_{13}^{\bf
k}|^2\, \sin^{2} \lf( \De_{13}^{\bf k} t \ri) + |V_{13}^{\bf
k}|^2\, \sin^{2} \lf( \Om_{13}^{\bf k}  t \ri)\Big]
\\ \non
& -&\cos^{2}\te_{13}\sin\te_{13}
 \Big[\cos\de\sin(2\te_{12})\sin(2\te_{23}) +
  4 \sin^2\te_{12}\sin\te_{13}\cos^2\te_{23}\Big]\Big[|U_{23}^{\bf k}|^2 \,
\sin^{2} \lf(\De_{23}^{\bf k} t \ri) +
 |V_{23}^{\bf k}|^2\,
\sin^{2} \lf(\Om_{23}^{\bf k}  t \ri)\Big]
\\ \non
&-& \cos^{2}\te_{13} \sin(2\te_{12}) \Big[ (\sin^2\te_{23} -
\sin^2\te_{13}\cos^2\te_{23})\sin(2\te_{12})
-\cos\de\cos(2\te_{12})\sin\te_{13}\sin(2\te_{23})\Big] \times
\\
&\times& \Big[|U_{12}^{\bf k}|^2\, \sin^{2} \lf( \De_{12}^{\bf k}
t \ri) + |V_{12}^{\bf k}|^2\, \sin^{2} \lf(\Om_{12}^{\bf k} t
\ri)\Big]\, . \eea

\section*{Appendix B: Useful formulas for the generation of the  mixing matrix }

In deriving the ${\cal U}_i$ mixing matrices of \S II and \S III,
we use the following relationships
\bea\
&&[\nu_{1}^{\al}(x),L_{12}]=\nu_{2}^{\al}(x)e^{-i\de_{12}}\;, \;
[\nu_{1}^{\al}(x),L_{23}]=0\qquad\qquad\quad, \;
[\nu_{1}^{\al}(x),L_{13}]=\nu_{3}^{\al}(x)e^{-i\de_{13}}\;,
\; \\
&&[\nu_{2}^{\al}(x),L_{12}]=-\nu_{1}^{\al}(x)e^{i\de_{12}}\;, \;
[\nu_{2}^{\al}(x),L_{23}]=\nu_{3}^{\al}(x)e^{-i\de_{23}}\;, \;
[\nu_{2}^{\al}(x),L_{13}]=0\;, \;
\\
&&[\nu_{3}^{\al}(x),L_{12}]=0\qquad\qquad\quad, \;
[\nu_{3}^{\al}(x),L_{23}]=-\nu_{2}^{\al}(x)e^{i\de_{23}}\;, \;
[\nu_{3}^{\al}(x),L_{13}]=- \nu_{1}^{\al}(x)e^{i\de_{13}}\;, \;
\eea and \bea\ &&G_{23}^{-1}(t)\nu_{1}^{\al}(x)G_{23}(t)=
\nu_{1}^{\al}(x)\;,\;
\\
&&G_{13}^{-1}(t)\nu_{1}^{\al}(x)G_{13}(t)=
\nu_{1}^{\al}(x)c_{13}+\nu_{3}^{\al}(x)e^{-i\de_{13}}s_{13}\;,\;
\\
&&G_{12}^{-1}(t)\nu_{1}^{\al}(x)G_{12}(t)=
\nu_{1}^{\al}(x)c_{12}+\nu_{2}^{\al}(x)e^{-i\de_{12}}s_{12}\;,\;
\eea

\bea &&G_{23}^{-1}(t)\nu_{2}^{\al}(x)G_{23}(t)=
\nu_{2}^{\al}(x)c_{23}+\nu_{3}^{\al}(x)e^{-i\de_{23}}s_{23}\;,\;
\\
&&G_{13}^{-1}(t)\nu_{2}^{\al}(x)G_{13}(t)= \nu_{2}^{\al}(x)\;,\;
\\
&&G_{12}^{-1}(t)\nu_{2}^{\al}(x)G_{12}(t)=
\nu_{2}^{\al}(x)c_{12}-\nu_{1}^{\al}(x)e^{i\de_{12}}s_{12}\;,\;
\eea

\bea &&G_{23}^{-1}(t)\nu_{3}^{\al}(x)G_{23}(t)=
\nu_{3}^{\al}(x)c_{23}-\nu_{2}^{\al}(x)e^{i\de_{23}}s_{23}\;,\;
\\
&&G_{13}^{-1}(t)\nu_{3}^{\al}(x)G_{13}(t)=
\nu_{3}^{\al}(x)c_{13}-\nu_{1}^{\al}(x)e^{i\de_{13}}s_{13}\;,\;
\\
&&G_{12}^{-1}(t)\nu_{3}^{\al}(x)G_{12}(t)= \nu_{3}\; \eea

\section*{Appendix C: Arbitrary mass parameterization and physical quantities}

In Ref. \cite{fujii1,fujii2} it was noticed that  expanding the
flavor fields in the same basis as the (free) fields with definite
masses (cf. Eq.(\ref{exnue1})) is actually a special choice, and
that a more general possibility exists. In other words, in the
expansion Eq.(\ref{exnue1}) one could  use eigenfunctions with
arbitrary masses $\mu_\sigma$, and therefore not necessarily the
same as the masses which appear in the Lagrangian.  On this basis,
the authors of Ref.\cite{fujii1,fujii2} have generalized the
Blasone-Vitiello (BV) formalism by writing the flavor fields as
\bea\label{exnuf2} \nu_{\sigma}(x)     &=& \sum_{r} \int d^3 k
\left[ u^{r}_{{\bf k},\sigma} {\widetilde \alpha}^{r}_{{\bf
k},\sigma}(t) + v^{r}_{-{\bf k},\sigma} {\widetilde
\beta}^{r\dag}_{-{\bf k},\sigma}(t) \right]  e^{i {\bf k}\cdot{\bf
x}} , \eea
where $u_{\sigma}$ and $v_{\sigma}$ are the helicity
eigenfunctions with mass $\mu_\sigma$. We denote by a tilde the
generalized flavor operators introduced in
Ref.\cite{fujii1,fujii2} in order to distinguish them from the
ones in BV formalism Eq.(\ref{exnue1}).  The expansion
Eq.(\ref{exnuf2}) is more general than the one in
Eq.(\ref{exnue1}) since the latter corresponds to the particular
choice $\mu_e\equiv m_1$, $\mu_\mu \equiv m_2$, $\mu_\tau\equiv
m_3$. Of course, the flavor fields in Eq.(\ref{exnuf2}) and
Eq.(\ref{exnue1}) are the same fields. The relation, given in
Ref.\cite{fujii1,fujii2}, between the general flavor operators and
the BV ones is
\bea\label{FHYBVa} &&\left(\begin{array}{c}
{\widetilde \alpha}^{r}_{{\bf k},\sigma}(t)\\
{\widetilde \beta}^{r\dag}_{{-\bf k},\sigma}(t)
\end{array}\right)
\;=\; J^{-1}_{\mu_{\sigma}}(t)  \left(\begin{array}{c}
\alpha^{r}_{{\bf k},\sigma}(t)\\ \beta^{r\dag}_{{-\bf
k},\sigma}(t)
\end{array}\right)J_{\mu_{\sigma}}(t) ~~,
\\ [2mm]\label{FHYBVb}
&&J_{\mu_{\sigma}}(t)\,=\, \prod_{{\bf k}, r}\, \exp\left\{ i
\mathop{\sum_{(\sigma,j)}} \xi_{\sigma,j}^{\bf k}\left[
\alpha^{r\dag}_{{\bf k},\sigma}(t)\beta^{r\dag}_{{-\bf
k},\sigma}(t) + \beta^{r}_{{-\bf k},\sigma}(t)\alpha^{r}_{{\bf
k},\sigma}(t) \right]\right\}\,, \eea
with $(\sigma,j)=(e,1) , (\mu,2), (\tau,3)$, $\xi_{\sigma,j}^{\bf
k}\equiv (\chi^{\bf k}_\sigma - \chi^{\bf k}_j)/2$ and
$\cot\chi^{\bf k}_\sigma = |{\bf k}|/\mu_\sigma$, $\cot\chi_j^{\bf
k} = |{\bf k}|/m_j$. For $\mu_\si\equiv m_j$, one has
$J_{\mu_{\sigma}}(t)=1$.

As already noticed in Ref.\cite{remarks}, the flavor charge
operators are the Casimir operators for the Bogoliubov
transformation (\ref{FHYBVa}), i.e. they are free from arbitrary
mass parameters : ${\wti Q}_\si(t) =Q_\si(t) $. This is obvious
also from the fact that they can be expressed in terms of flavor
fields (see Ref.\cite{comment}).

Physical quantities should not carry any dependence on the
$\mu_\si$: in the two--flavor case, it has been shown
\cite{remarks} that the expectation values of the flavor charges
on the neutrino states are free from the arbitrariness. For three
generations, the question is more subtle due to the presence of
the CP violating phase. Indeed, in Ref.\cite{fujii2} it has been
found that the corresponding generalized quantities depend on the
arbitrary mass parameters.

In order to understand better the nature of such a dependence, we
consider the identity:
\bea\lab{gen1}
 \langle {\ti \psi}|{\ti Q}_\si(t)| {\ti
\psi}\ran \, =\, \langle \psi| J(0) \, Q_\si(t)\, J^{-1}(0)|
\psi\ran \, =\, \langle \psi|  Q_\si(t) | \psi\ran + \langle \psi|
\lf[J(0) , Q_\si(t)\ri]\, J^{-1}(0)|  \psi\ran\,. \eea
valid on any vector $|\psi\ran$ of the flavor Hilbert space (at
$t=0$). From the explicit expression for $J(0)$ we see that the
commutator $\lf[J(0) , Q_\si(t)\ri]$ vanishes for $\mu_\rho=m_j$ ,
$(\rho,j)=(e,1) , (\mu,2), (\tau,3)$.

It is thus tempting to define the (effective) physical flavor
charges as:
\bea {\wti {Q}}^{phys}_{\si}(t)\, \equiv \,{Q}_{\si}(t) -
J^{-1}(0)\lf[J(0) , Q_\si(t)\ri]\,= \, J^{-1}(0) \, {Q}_{\si}(t)\,
J(0),
 \eea
such that for example:
\bea \langle {\ti \nu_\rho}|{\wti {Q}}^{phys}_{\si}(t) \,| {\ti
\nu_\rho}\ran \, =\,\langle \nu_\rho| Q_\si(t) |  \nu_\rho\ran.
 \eea

It is clear that the operator ${\wti {Q}}^{phys}_{\si}(t)$ does
depend on the arbitrary mass parameters and this dependence is
such to compensate the one arising from the flavor states. The
choice of physical quantities (flavor observables) as those not
depending on the arbitrary mass parameters is here adopted,
although different possibilities are explored by other authors,
see Refs.\cite{fujii2,Ji2,Ji3}.

\end{document}